\documentclass[twocolumn]{aastex631_1}

\usepackage{newtxtext,newtxmath}
\usepackage[T1]{fontenc}
\usepackage{ae,aecompl}
\usepackage{multirow}
\usepackage{xcolor}
\usepackage{mathtools}
\usepackage{anyfontsize}
\usepackage{graphicx}	
\usepackage{amsmath}	
\newcommand{\bz}{$\langle B_z \rangle$}

\newcommand{\ra}{$R_\mathrm{A}$}
\newcommand{\rk}{$R_\mathrm{K}$}

\begin{document}
\nolinenumbers
\title{Coherent radio emission from `Main-sequence Radio Pulse emitters': a new stellar diagnostic to probe 3D magnetospheric structures}

\author[0000-0001-8704-1822]{Barnali Das}
\affiliation{CSIRO, Space and Astronomy, P.O. Box 1130, Bentley WA 6102, Australia}

\author[0000-0002-0844-6563]{Poonam Chandra}
\affiliation{National Radio Astronomy Observatory, 520 Edgemont Road, Charlottesville VA 22903, USA}

\author[0000-0002-5633-7548]{V\'eronique Petit}
\affiliation{Department of Physics and Astronomy, Bartol Research Institute, University of Delaware, 104 The Green, Newark, DE 19716, USA}



\begin{abstract}
Main-sequence Radio Pulse emitters (MRPs) are magnetic early-type stars that produce coherent radio emission observed in the form of periodic radio pulses. The emission mechanism behind is the Electron Cyclotron Maser Emission (ECME). 
Amongst all kinds of magnetospheric emission, ECME is unique due to its high directivity and intrinsically narrow bandwidth.
The emission is also highly circularly polarized and the sign of polarization is opposite for the two magnetic hemispheres. This combination of properties makes ECME highly sensitive to the three-dimensional structures in the stellar magnetospheres. 
This is especially significant for late-B and A-type magnetic stars that do not emit other types of magnetospheric emission such as H$\alpha$, the key probe used to trace magnetospheric densities.
In this paper, we use ultra-wideband observation (0.4-2 GHz) of a late B-type MRP HD\,133880 to demonstrate how we can extract information on plasma distribution from ECME. We achieve this by examining the differences in pulse arrival times (`lags') as a function of frequencies, 
and qualitatively comparing those with lags obtained by simulating ECME ray paths in hot stars' magnetospheres.
This reveals that the stellar magnetosphere has a disk-like overdensity inclined to the magnetic equator with a centrally concentrated density that primarily affects the intermediate frequencies (400--800 MHz).  
This result, which is consistent with recent density model proposed for hotter centrifugally supported magnetospheres, lends support to the idea of a unifying model for magnetospheric operations in early-type stars, and also provides further motivation to fully characterize the ECME phenomenon in large-scale stellar magnetospheres.
\end{abstract}



\section{Introduction} \label{sec:intro}
Magnetic early-type (spectral types OBA) stars constitute approximately 10\% of the early-type star population \citep[e.g.][]{grunhut2017}. They are unique objects in the stellar main-sequence due to their unusually stable surface magnetic fields that can often be approximated as axi-symmetric dipoles inclined to their rotation axes \citep[][etc.]{shultz2018}. 
The stability and simplicity of these magnetic fields make their hosts ideal celestial laboratories to understand the physics of large-scale stellar magnetospheres. Such magnetospheres form due to the interaction of the radiatively driven strong stellar wind of the hot stars with their kG-strength magnetic fields \citep[e.g.][]{andre1988,trigilio2004,townsend2005}.

In the past, several studies have been performed to understand the formation and operation of magnetospheres surrounding O and early-B type stars \citep[e.g.][]{petit2013,naze2014,shultz2018,shultz2019b,shultz2019c,shultz2020}. While these stars produce emission over a wide range of the electromagnetic spectrum (from X-ray to radio), the emission that has been most extensively used to probe their magnetospheres is the H$\alpha$ emission. 
H$\alpha$ is produced by clouds of co-rotating magnetospheric plasma (regions with high plasma densities), and the line-profile shape and strength contain information regarding the location of the plasma clouds as well as their densities \citep[e.g.][]{townsend2005,oksala2015,shultz2020,owocki2020}. In other words,
the rotational modulation of the spectral line profile provides
key information about the three-dimensional magnetospheric structures \citep{townsend2005}. Studies involving H$\alpha$ emission led to the development of several fundamental concepts such as the classification based on the relative values of the Kepler radius \rk~and the Alfv\'en radius \ra. Kepler radius is the distance at which centrifugal force due to co-rotation balances gravity (thus, \rk~increases with increasing rotation period). Alfv\'en radius, on the other hand, defines the extent of the magnetosphere, beyond \ra, the wind kinetic energy is stronger than the magnetic field energy, and the field lines are drawn open. Stars that have strong magnetic fields (hence large \ra) and also are fast-rotating (resulting in smaller \rk)
can have \ra~larger than \rk. In those cases, there is a region between \rk~and \ra, named as `centrifugal magnetosphere' (CM), where stellar wind matter can accumulate up to very high density at certain locations where gravitational force, centrifugal force and magnetic field tension balance one another. This leads to the formation of disk-like structures in the CM that manifest themselves through their effects on various radiation (originating from the stellar surface and from the magnetosphere) that pass through those regions \citep{townsend2005}.
These high-density regions are believed to be stable structures even though there is a continuous injection of wind materials. This is achieved via `centrifugal breakout' (CBO), small spatial-scale magnetospheric explosions that happen at all times facilitating ejection of excess plasma away from the CM \citep{shultz2020,owocki2020}. Most recently, it has been shown that CBOs also drive incoherent radio emission by providing the non-thermal electrons through magnetic reconnections \citep{leto2021,shultz2022,owocki2022}.

Despite the success of H$\alpha$ emission, it has the limitation that the emission is absent in late-B and A-type stars raising the possibility that the cooler magnetic stars with much weaker winds may not operate in the same way as their hotter counterparts do \citep{shultz2020}.
However, \citet{leto2021} and \citet{shultz2022} showed that the incoherent radio emission does not exhibit a dependence on the effective temperature and the emission can be described in the CBO-framework for all magnetic early-type stars with CM (not considering binarity). 
The importance of magnetospheric radio emission (as a unifying characteristic for large-scale stellar magnetospheres) becomes much more prominent when we consider the similarities of magnetic early-type stars with much cooler, magnetic ultracool/brown dwarfs. Initially, the similarities were noted only in their coherent radio emission, but in recent times, striking similarities were reported in the characteristics of their incoherent radio emission as well \citep[][etc.]{trigilio2000,berger2001,leto2021,callingham2023,bloot2024}. 
\citet{leto2021} showed that both hot and cool magnetic stars appear to follow the same scaling law connecting the incoherent radio luminosity and stellar parameters;
\citet{bloot2024} reported rotational modulation of circular polarization fraction of incoherent radio emission from a magnetic cool star, also exhibited by magnetic hot stars \citep[][etc.]{lim1996,leto2017}. All these evidences suggest that the physics of large-scale magnetospheres might be the same across various spectral types.

To be able to use the magnetic hot stars as true celestial laboratories for investigating magnetospheric physics that might have applications beyond the OBA spectral types, it is imperative that we understand them as best as possible. 
Though significant progress has already been made along that direction, with incoherent radio emission acting as the primary probe for cooler magnetic early-type stars, the latter is not as sensitive to the three dimensional structures as the H$\alpha$ emission, and is more suitable to estimate average quantities \citep[e.g.][]{trigilio2004,leto2006,leto2021}. 
Another disadvantage of currently used magnetospheric probes including H$\alpha$ and incoherent radio emission is that they all originate at large emission sites, so that they cannot probe events that occur at small spatial scales (e.g. direct evidence of CBOs). With these probes, it is also challenging to characterize how deviations from simplifying assumptions made in theories (such as assumption of an axi-symmetric dipole) affect the magnetospheric properties.
In this paper, we propose that \textit{coherent} radio emission, observed as periodic radio pulses and speculated to be ubiquitous among magnetic massive stars \citep{das2022a}, provides us with a probe that has the potential to complement the existing set of magnetospheric probes by overcoming their limitations. 
At the moment, our understanding about the phenomenon (coherent radio emission) is incomplete so that we only focus on extracting qualitative information. Thus, a secondary purpose of this paper is to highlight the need to develop a more detailed understanding about coherent radio emission from magnetic hot stars, considering the enormous potential it holds to become a unique magnetospheric probe.

This paper is structured as follows: in the next section (\S\ref{sec:strategy}), we describe the principle behind the methodology adopted to acquire magnetospheric information, and a brief description of our strategy, this is followed by a description of the observations used (\S\ref{sec:obs}). We then describe how we use the wideband observations to measure difference in pulse arrival phases (called `lags') between a pair of frequencies (\S\ref{sec:finding_lags}) and present the spectral variation of the lags in \S\ref{sec:results}. These lags are our key observables to extract information about the three dimensional magnetosphere.
We interpret the results in \S\ref{sec:significance} and then conclude the paper with a discussion of the results in \S\ref{sec:summary}.

\section{Principle and Strategy}\label{sec:strategy}
\begin{figure*}
    \centering
    \includegraphics[width=0.85\textwidth]{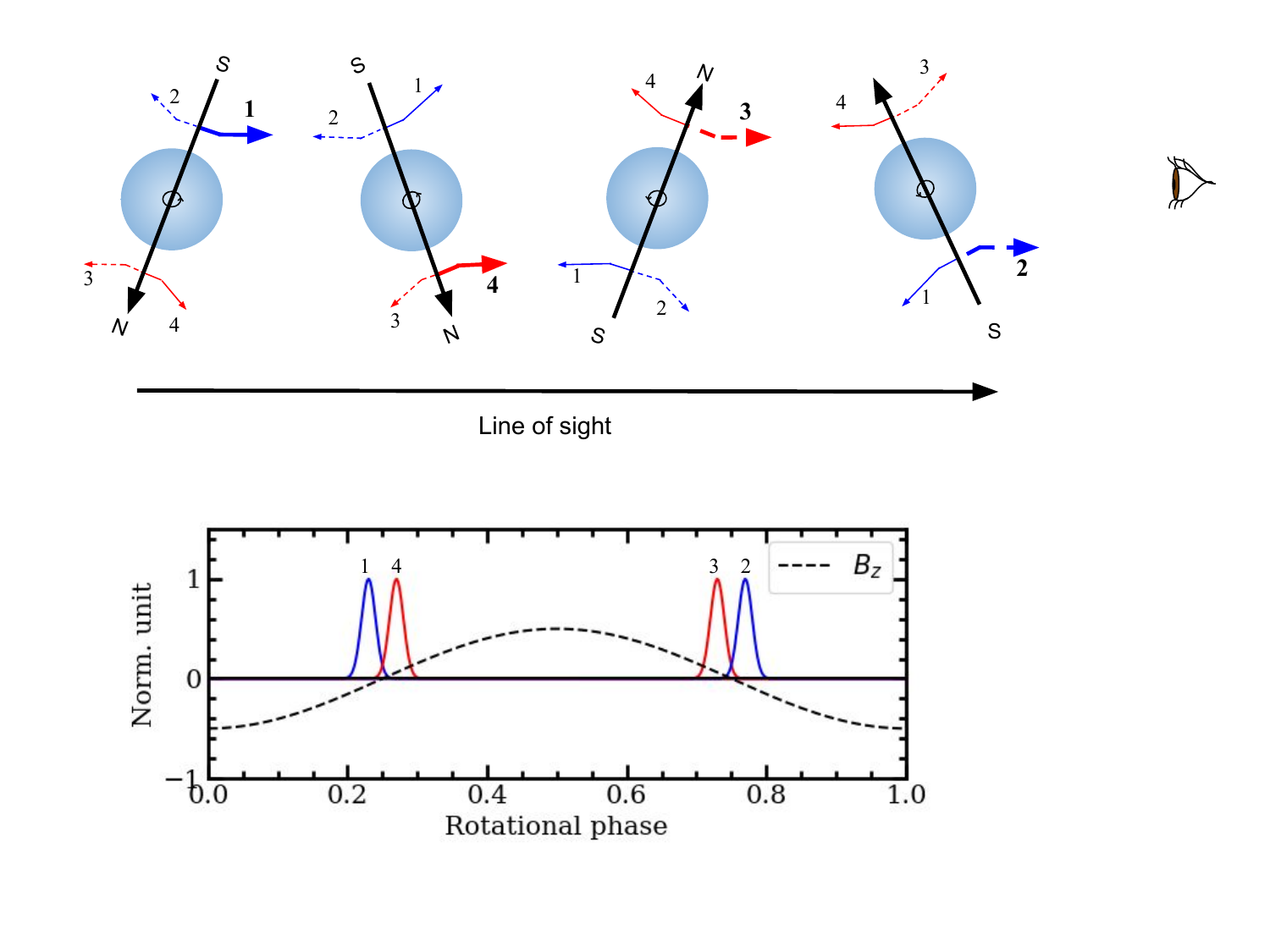}
    \caption{A cartoon diagram demonstrating how ECME from a star with an axi-symmetric dipolar magnetic field conveys information regarding different parts of the stellar magnetosphere. For simplicity, we have assumed that the line of sight and the rotation axes are perpendicular to each other, and the magnetic axis is also perpendicular to the rotation axis. Rotation axis is perpendicular to the page. Intrinsically, the radiation is nearly at right angle to the magnetic axis, but get deviated due to propagation effects in the magnetosphere (indicated by bent arrows). In one full rotation cycle, two pairs of pulses are observed around the rotational phases where the longitudinal magnetic field \bz~is zero (called magnetic nulls). Each pair has one pulse coming from the northern and one coming from the southern magnetic hemispheres that can be distinguished based on their opposite circular polarizations. Note that pulses of same polarizations, but observed around different magnetic nulls, trace different parts of the stellar magnetospheres. The rays are labelled with numbers to indicate the pulses that they give rise to in the lightcurves.}
    \label{fig:ecme_ideal}
\end{figure*}
Coherent radio emission from hot magnetic stars was first discovered by \citep{trigilio2000}. 
The emission was distinguishable from incoherent gyrosynchrotron emission because it is nearly 100\% circularly polarized, has much higher flux density, and is highly directed. The latter was inferred because the emission was observed as pulses at fixed rotational phases of the star, similar to the case of pulsars.
Based on these properties, the mechanism was identified to be electron cyclotron maser emission \citep[ECME,][]{trigilio2000,trigilio2008}. ECME can operate when there is an unstable distribution of electrons (such as loss-cone distribution) gyrating in a magnetic field, provided the local plasma frequency is less than the electron gyrofrequency \citep{melrose1982,treumann2006}. It is an intrinsically narrow-band phenomenon with the frequency of emission being proportional to the electron gyrofrequency at the site of emission, and thus proportional to the local magnetic field strength. As a result, higher frequencies are produced closer to the star and vice-versa. 

\citet{das2021} introduced the name of `Main-sequence Radio Pulse emitters' (MRPs) to describe the magnetic hot stars producing ECME. The current number of known MRPs is 18 \citep{trigilio2000,chandra2015,das2018,leto2019,lenc2018,das2019a,das2019b,leto2020,leto2020b,pritchard2021,das2022a,das2022b}.

Observations of ECME from the first discovered MRP CU\,Virginis (hereafter CU\,Vir) suggested that it is produced near the stellar magnetic poles in ring-shaped regions, called `auroral rings' \citep{trigilio2011}. Depending on whether the radiation is produced in the extra-ordinary (X-) or ordinary (O-) mode, the north magnetic pole produces right (RCP) or left circularly polarized (LCP) radiation, respectively, and vice-versa \citep[e.g., see Figure 6 of][]{leto2016}. Radiation is directed tangential to the auroral rings in such a way that it is perpendicular to the local magnetic field direction and parallel to the magnetic equator \citep{trigilio2011}. This beaming model, referred as `tangent plane beaming model', was proposed to explain the appearance of the radio pulse from CU\,Vir near the rotational phases where the stellar surface averaged line of sight magnetic field (longitudinal magnetic field \bz) is zero. These rotational phases are called magnetic nulls.

Once the radiation is produced, it has to pass through the magnetosphere to reach the observer \citep[see Figure 2 of ][]{leto2016}. The presence of the confined plasma inside the magnetosphere makes the radiation deviate from their original direction of emission \citep{trigilio2011}. The deviation suffered is a function of frequency and plasma density. As a result, the frequency dependence of the pulse arrival time contains information about the inner magnetospheric plasma. Since LCP and RCP pulses pass through different parts of the inner magnetosphere, their observation over wide frequency bands allows us to compare the northern and southern parts of the stellar magnetosphere. This is illustrated in Figure \ref{fig:ecme_ideal}.

\citet{das2021} proposed that the unique combination of properties makes ECME capable of sensing CBOs. Here, we will focus on how one can extract information about the magnetospheric plasma distribution with the help of wideband observations.
The key idea was first proposed by \citet{das2020a0}, where it was shown that over a small range of frequencies, the lag between a pair of frequencies will approximately vary as the difference between the square of their wavelengths. The constant of proportionality will contain information regarding the average plasma density and by measuring that constant, the average density can be estimated. This is similar to using dispersion measures to infer densities, except that here the refractions occur in the magnetized plasma of the stellar magnetosphere.
Subsequently, \citet{das2020b,das2020a} suggested that ECME properties are also sensitive to the three dimensional plasma density distribution, and that ECME could become a probe for tracing that distribution. The work presented in this paper is the first step towards implementing this idea.

In this paper, we use wideband observation of ECME from the hot magnetic star HD\,133880. We measure the differences between the rotational phases of arrival of pulses of a pair of frequencies for one of the RCP and one of the LCP pulses. The results obtained for the two polarizations are compared. We then used the framework presented in \citet{das2020b} to understand the significance of the difference exhibited by the pulses produced by opposite magnetic polar regions.

\begin{deluxetable*}{cccccc}
\tabletypesize{\scriptsize}
\centering
\tablecaption{Observation details of the data used in this work. The rotational phases ($\Delta \phi_\mathrm{rot}$) are calculated using the ephemeris in \citet{das2018}. All but the data acquired on 2019--05--03 (boldfaced) were reported by \citet{das2020b}.\label{tab:targets_obs}}  
\tablehead{
Telescope & Date & $\Delta \phi_\mathrm{rot}$ & Eff. band & \multicolumn{2}{c}{Calibrator}\\
(band name) & of Obs. & & $\Delta\nu_\mathrm{eff}$ (MHz) & Flux/bandpass & Phase 
}
\startdata
\hline
uGMRT & 2019--05--17 & 0.10--0.35 & 334--461 & 3C286 & J1517--243, J1626--298  \\
(band 3) & 2019--03--17 &  0.63--0.88 & 334--360, 380--461 & 3C286 & J1517--243, J1626--298 \\
\hline
uGMRT & \textbf{2019--05--03} & \textbf{0.19--0.25} & \textbf{570--804} & \textbf{3C286} & \textbf{J1626--298} \\
(band 4) & 2019--08--02 & 0.61--0.81 &  570--804 & 3C286 & J1517--243, J1626--298\\
\hline
VLA & 2019--09--07 & 0.13--0.30 & 1040--1104, 1360--1488, 1680--2000 & 3C286 & J1522--2730 \\
(L)& 2019--11--01 & 0.63--0.80 & 1040--1104, 1360--1488, 1680--2000 & 3C286 & J1522--2730 \\
\enddata
\end{deluxetable*}

\section{Observation and data reduction}\label{sec:obs}
We observed the late B-type star HD\,133880 \citep{buscombe1969} over a frequency range of 0.4--4 GHz using two radio telescopes: the upgraded Giant Metrewave Radio telescope (uGMRT, 384--800 MHz) and the Karl G. Jansky Very Large Array (VLA, 1--4 GHz). All the data was acquired in the year 2019. The observations were scheduled so as to observe the star around the rotational phases corresponding to the nulls of its \bz~(where we expect to see ECME pulses). The ephemeris used to phase the data is that used in \citet{das2018}. According to this ephemeris, the two magnetic nulls correspond to rotational phases of 0.175 cycle (\bz~changes from negative to positive) and 0.725 rotation cycle (\bz~changes from positive to negative) \citep{das2020b}. The rotational phase 0.175 will be referred as `Null 1' and the other magnetic null phase (0.725 rotation cycle) will be called `Null 2'.

The uGMRT data over $\approx 400-800$ MHz were acquired by observing over two bands: band 3 (300--500 MHz) and band 4 (550--900 MHz). All the uGMRT data in band 3, which are used in this work, are already described in \citet{das2020b}. The data in band 4 near Null 2 have also been reported in \citet{das2020b}. But around Null 1 (band 4), we do not use the corresponding data described in \citet{das2020b} since those data were acquired in the year 2018 and thus have a significant gap with rest of the data.
The requirement of near-simultaneity and the effect of using non-simultaneous data are explained in \S\ref{sec:non-simultaneity}. Near null 1, we had observed the star in band 4 on 2019 May 3, but unfortunately these data only covered the RCP pulses. We will hence restrict our analysis to the RCP pulses for null 1. 


The VLA data used here are the same as those used in \citet{das2020b}. They cover the frequency range 1--4 GHz in a continuous fashion distributed over two bands: L (1--2 GHz) and S (2--4 GHz). The L band is divided into 16 spectral windows (spws) each of width 64 MHz, and the S band is also divided into 16 spws each of which covers a frequency range of 128 MHz.  For more details on these data (as well as the uGMRT data) and the analysis procedure, please refer to the `Observation and Data analysis' section of \citet{das2020b}.

The details of the data used in this work are given in Table \ref{tab:targets_obs}. 
The data are phased assuming a constant rotation period $P_\mathrm{rot}=0.877483$ days \citep{kochukhov2017} and a reference heliocentric Julian day (HJD) of 2445472.00 \citep[used by][]{das2018}.

\section{`Lag' between ECME pulses at two different frequencies}\label{sec:finding_lags}
In order to obtain the `lag' (denoted by $\tau$), i.e. the difference between the rotational phase of arrival of pulses at two different frequencies, we choose to work with frequencies separated by $\approx 200$ MHz spanning 400--2000 MHz. Above 2 GHz, ECME pulses are very weak \citep[the peak flux density declines as a power-law with a slope of $\approx -2$ above 1 GHz, also the upper cut-off frequency for the RCP pulse lies at $\approx 3.2$ GHz, see Figures 6 and 7, and \S3.5 of][]{das2020b} so that the pulses are not well-sampled (due to low SNR), hence the frequency range above 2 GHz will not be used in this work.

The effective bandwidth of different wavebands are shown in the fourth column of Table \ref{tab:targets_obs}. As described in \citet{das2020b}, uGMRT band 3 and band 4 were further divided into smaller spectral bins in order to investigate the intra-band spectral behaviour. From band 3, the lightcurves at the spectral bin centred at 398/395 MHz (closest to 400 MHz), with a bandwidth of $\approx 25$ MHz are chosen. From band 4, two spectral bins (each of width $\approx 47$ MHz), centred at 593 MHz (closest to 600 MHz) and 781 MHz (closest to 800 MHz) are chosen. Finally, from the VLA L-band that is divided into 16 spectral windows (spws), we choose the spws centred at 1040 MHz (closest to 1 GHz), 1104 MHz (closest to 1.2 GHz), 1424 MHz (closest to 1.4 MHz), 1680 MHz (closest to 1.6 GHz), 1808 MHz (closest to 1.8 GHz) and 1999.5 MHz (closest to 2.0 GHz). The highest spectral bin is discarded as the pulses were not adequately sampled (due to low SNR).

\begin{figure}
    \centering
    \includegraphics[width=0.49\textwidth]{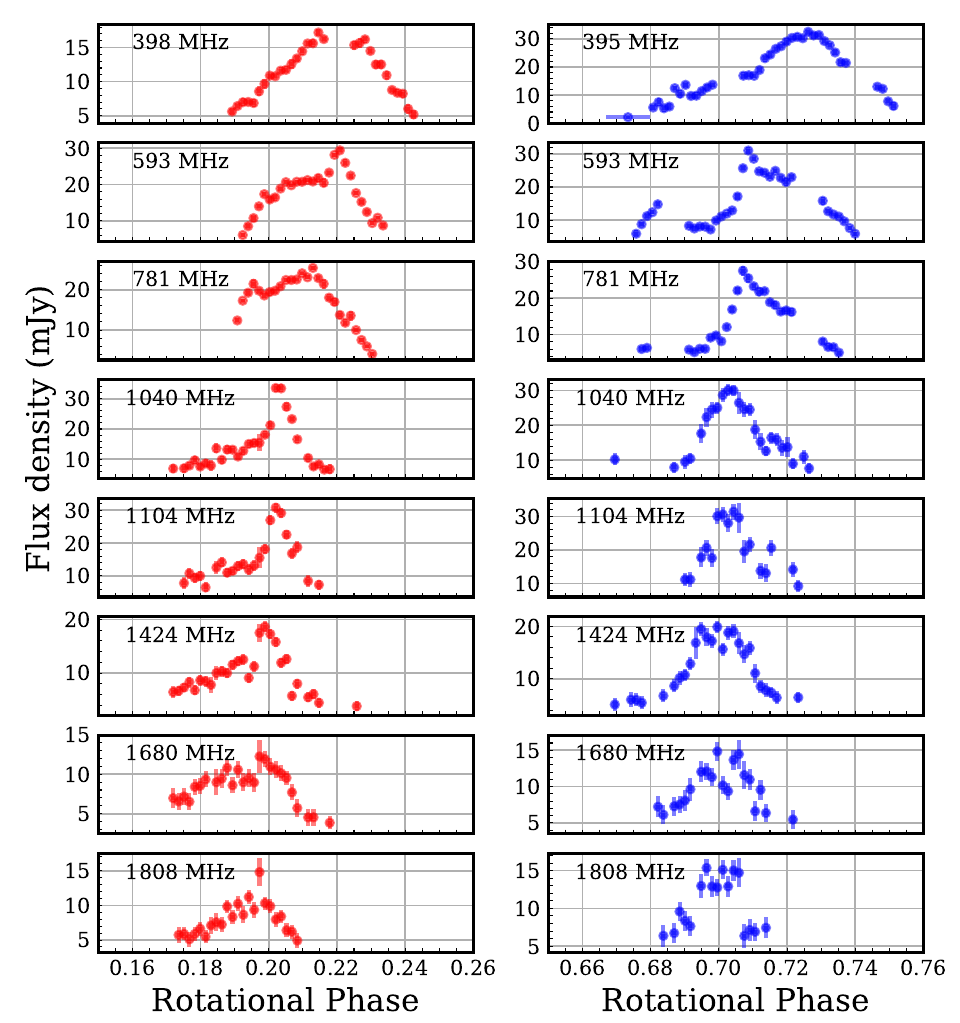}
    \caption{The RCP (red markers, left panels) and LCP (blue markers, right panels) ECME pulses from HD\,133880 near null 1 (left panels) and null 2 (right panels) over 400--2000 MHz. One can see that the pulses at different frequencies are shifted in rotational phases w.r.t. one another.}
    \label{fig:hd133880_lightcurves}
\end{figure}

Near each null, there is a pair of RCP and LCP pulses (see Figure \ref{fig:ecme_ideal}). However, as mentioned already, for the data taken on 2019 May 3 (at band 4 near null 1), we do not have coverage for the LCP pulse. Hence we consider only the RCP pulses near null 1 for obtaining the lags. In case of null 2, the RCP pulse was not detected in band 3 and very weak in band 4 \citep{das2020b}. As a result, we will consider only the LCP pulses near null 2. The lightcurves that we use in the subsequent exercise are shown in Figure \ref{fig:hd133880_lightcurves}.

The lags between the different frequencies are obtained by cross-correlation. The errorbars in the lags are obtained adopting a Monte Carlo approach (details in \S\ref{subsec:lag_lag_err}). The different steps involved in obtaining the lags and their errorbars are described below.

\subsection{Regridding and rescaling the individual lightcurves}\label{subsec:smooth_pad_lightcurves}
From Figure \ref{fig:hd133880_lightcurves}, we can clearly see that the pulses at different frequencies have different profiles and heights. This can artificially increase the errorbars in the lags. To suppress the effects, we perform a number of operations before cross-correlating the lightcurves.

For a given polarization (equivalently, a given magnetic null), we use a common phase-axis ($\phi_0$, uniform grid) to interpolate the flux density values at all frequencies (explained further in the next paragraph). 
To overcome the limitation of different pulse-heights (peak flux densities) at different frequencies, we perform a linear transformation on the flux densities at a given frequency and polarization. If $S_0$ is the original flux density at a particular polarization and at frequency $\nu$ at a given rotational phase, the new flux density after the linear transformation will be $\Tilde{S}=(S_0-S_\mathrm{min})/(S_\mathrm{max}-S_\mathrm{min})$; where $S_\mathrm{max}$ and $S_\mathrm{min}$ are respectively the maximum and minimum flux density values of the lightcurve at that polarization and at frequency $\nu$. The result of this transformation is that it maps all the values to lie between 0 and 1.

\subsection{Obtaining the lags $\tau$ and errorbars using Monte Carlo approach}\label{subsec:lag_lag_err}
Once we have the lightcurves (at a given circular polarization) all of which have flux density values lying between 0 and 1, over a common phase-axis,
we compute the value of the cross-correlation for each value of lag $\tau$ implied by the $\phi_0$ array. For each pair of frequencies we select the value of $\tau$ ($\tau_0$) that maximizes the cross-corelation. 

\begin{figure}
    \centering
    \includegraphics[trim={0cm 0cm 15cm 0cm}, clip, width=0.4\textwidth]{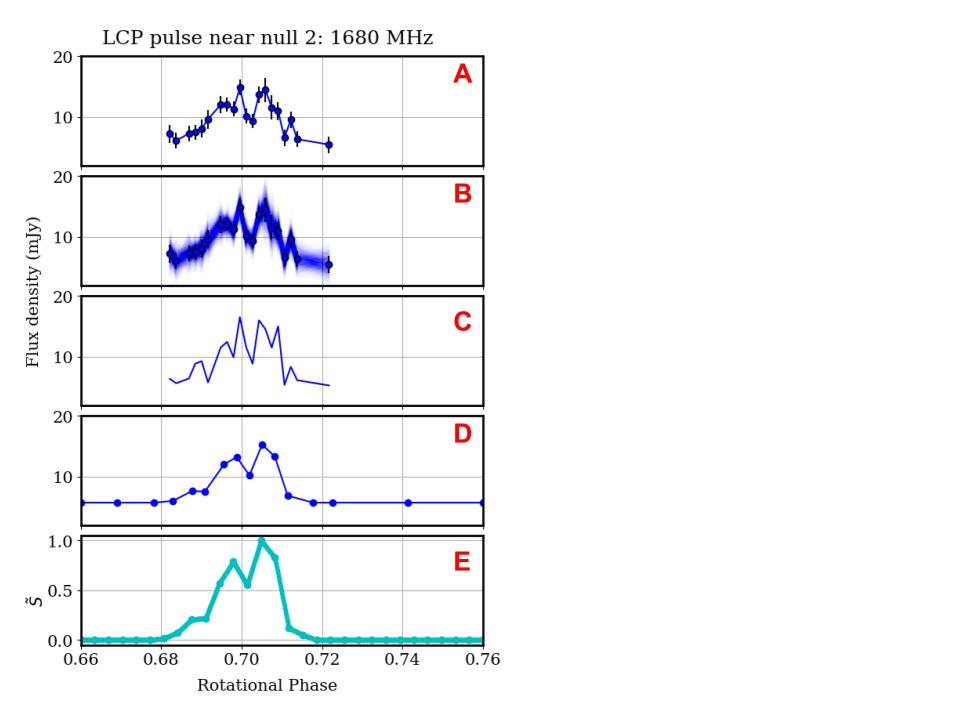}
    \caption{Demonstrating how the lightcurve at a given frequency and polarization is treated before we use it for cross-correlation. The top panel (marked `\textbf{A}') shows the observed lightcurve.
    The second panel (\textbf{B}) shows the $N$ (taken to be 2000 in the figure) lightcurves obtained by drawing $N$ random numbers for each data point assuming a Gaussian distribution with mean and sigma equal to the observed flux densities and errorbars respectively. Also shown are the actual data points (in blue circles). The third panel (\textbf{C}) shows one of the $N$ such lightcurves. The fourth panel (\textbf{D}) shows the lightcurve after it is smoothed and padded to span the chosen rotational phase range (0.66--0.76 in this case). The bottom panel (\textbf{E}) shows the lightcurve after it is resampled on a common rotational phase axis $\phi_0$ via interpolation, and the flux densities are linearly transformed to lie between 0 and 1.
     See \S\ref{subsec:smooth_pad_lightcurves} for the definition of $\tilde{S}$.  
    }
    \label{fig:LCP_demonstration}
\end{figure}

To compute the uncertainty on $\tau_o$, we employ the following Monte Carlo procedure (demonstrated in Figure \ref{fig:LCP_demonstration}):

For a given polarization,
\begin{enumerate}
    \item Let the observed flux densities at frequency $\nu_1$ be $S_\mathrm{obs}$ with errorbars $S_\mathrm{obs}^\mathrm{err}$ at rotational phases $\phi_\mathrm{obs}$. For each data point at phase $\phi_\mathrm{obs,0}$ on the lightcurve, random numbers are drawn from a Gaussian distribution with mean equal $S_\mathrm{obs,0}$ and standard deviation equal $S_\mathrm{obs,0}^\mathrm{err}$. This was done $N=20000$ times. After this exercise, we have $N$ number of lightcurves at each frequency (panel \textbf{B} in Figure \ref{fig:LCP_demonstration}).
    \item Each of these $N$ lightcurves are smoothed by convolving with a box-function that performs averaging over two consecutive data points.
    \item These `N' lightcurves are resampled over the $\phi_0$ grid using interpolation. In case, the $\phi_0$ grid extends beyond the $\phi_\mathrm{obs}$ grid, we pad the array of flux densities with the minimum of the two edge values before performing the resampling (panel \textbf{D} in Figure \ref{fig:LCP_demonstration}). In addition, we perform linear transformation so that the flux densities lie between 0 and 1 (panel \textbf{E} in Figure \ref{fig:LCP_demonstration}). 
    \item These lightcurves are now ready for cross-correlating with the corresponding lightcurve at a different frequency $\nu_2$. However, to reduce the step-size of `sliding' during cross-correlation of two lightcurves\footnote{While cross-correlating two arrays, we effectively slide one array over the other; multiplying the respective elements of the arrays at each step of sliding}, we now consider an array of rotational phases over the same range as $\phi_0$ (see \S\ref{subsec:smooth_pad_lightcurves}), but with a much higher resolution. Let us call this denser array of rotational phases to be $\phi_\mathrm{rot}$. We use $\phi_\mathrm{rot}$ to calculate the lags. Flux densities at $\phi_\mathrm{rot}$ are obtained by interpolation.
    \item These lightcurves with a common phase axis ($\phi_\mathrm{rot}$) and a common range of flux densities (0--1) are used to obtain lags via cross-correlation. For each pair of frequencies, $N=20000$ lags are obtained.
    \item For this lag distribution, we obtain a histogram where each bin corresponds to one element of lag $\tau$. The element in $\tau$ which has the highest frequency of occurrence is taken to be the value of lag for the pair of frequencies under consideration.
    \item From the histogram of the lag distribution, we calculate the cumulative distribution. From that, we obtain the interval in $\tau$ within which 90\% of the values lie. This interval is assigned as the uncertainty for the lag estimated for the given pair of frequencies and at the given circular polarization.
\end{enumerate}

\begin{figure}
    \centering
    \includegraphics[width=0.3\textwidth]{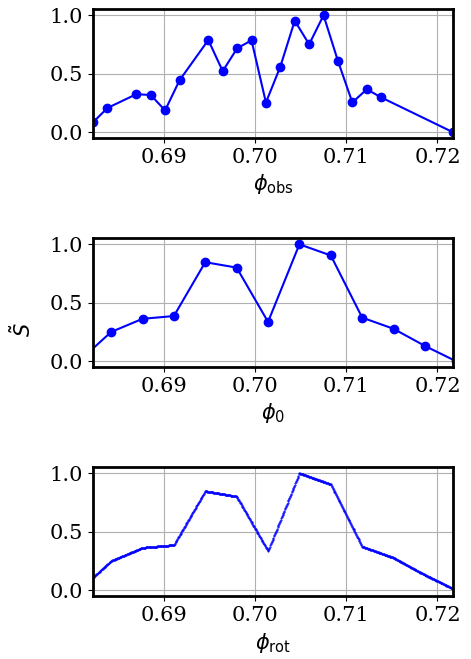}
    \caption{The difference among the different phase grids: $\phi_\mathrm{obs}$ (top) corresponds to the rotational phases in the `raw' lightcurves (values of rotational phases where flux densities are measured); $\phi_0$ (middle) is a uniform grid over a chosen phase range common to all the frequencies. Flux densities at $\phi_0$ are obtained from the measured values via interpolation; $\phi_\mathrm{rot}$ (bottom) is also a uniform grid over the same range as that spanned by $\phi_0$, but with a much higher resolution.
    }
    \label{fig:phase-grid_demonstration}
\end{figure}

The three phase-grids $\phi_\mathrm{obs}$, $\phi_0$ and $\phi_\mathrm{rot}$ are illustrated in Figure \ref{fig:phase-grid_demonstration}.

\section{Results}\label{sec:results}
\begin{figure}
    \centering
    \includegraphics[width=0.49\textwidth]{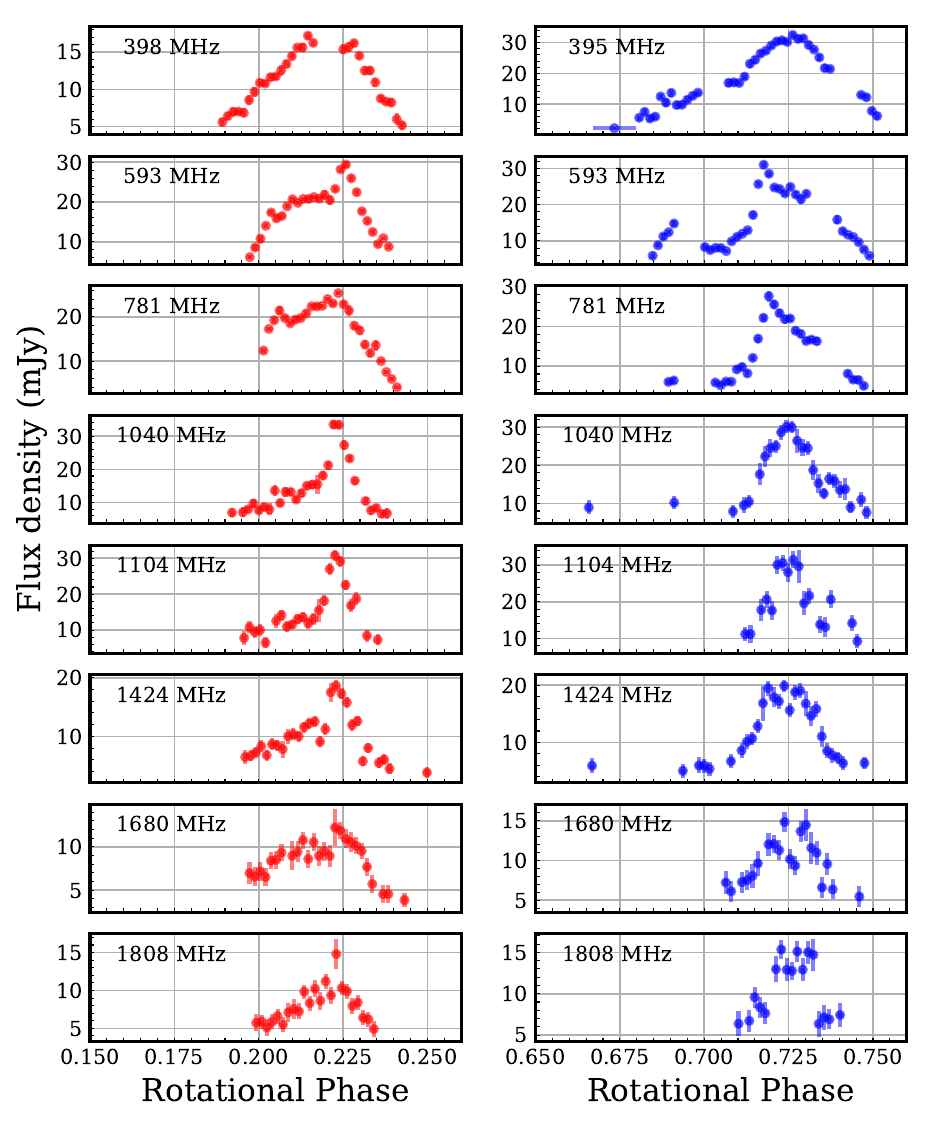}
    \caption{Same as Figure \ref{fig:hd133880_lightcurves}, but the lightcurves are now aligned to the lightcurves at the lowest frequency using the lags obtained by the procedure described in \S\ref{sec:finding_lags}.}
    \label{fig:hd133880_lightcurves_aligned}
\end{figure}
\begin{figure*}
    \centering
    \includegraphics[width=0.45\textwidth]{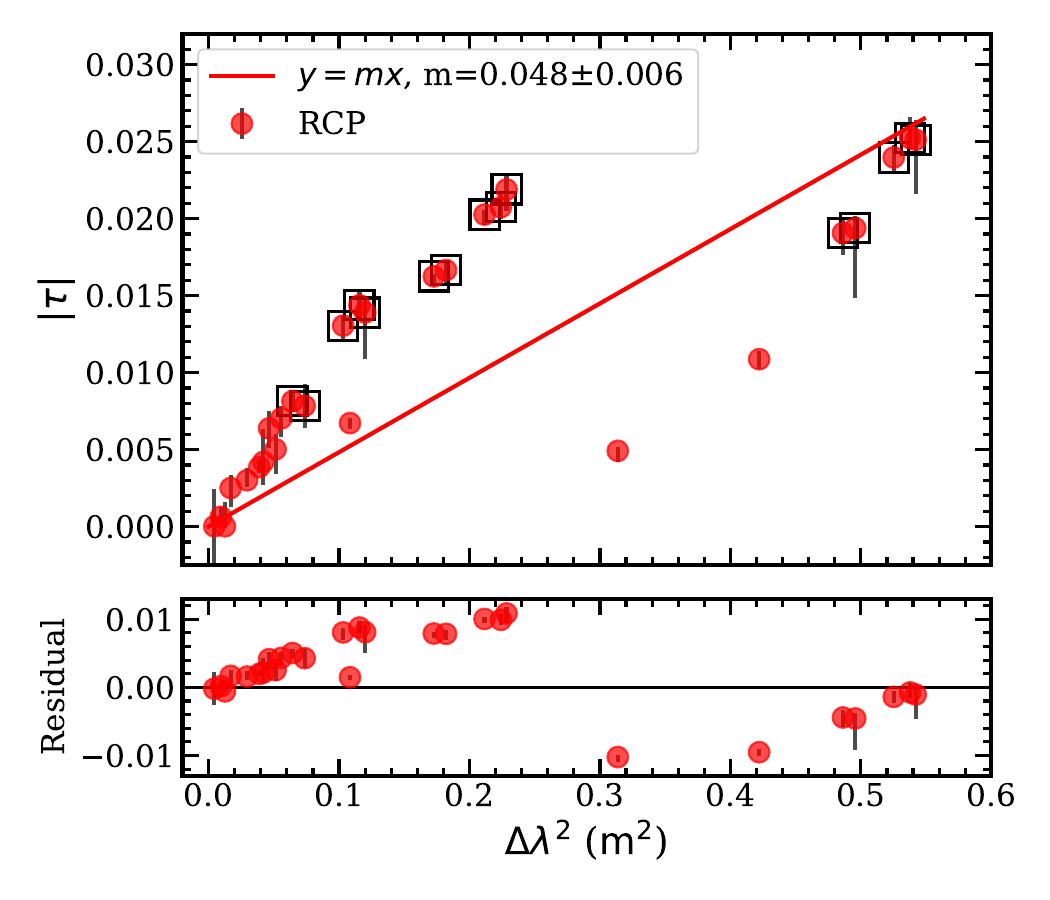}
    \includegraphics[width=0.45\textwidth]{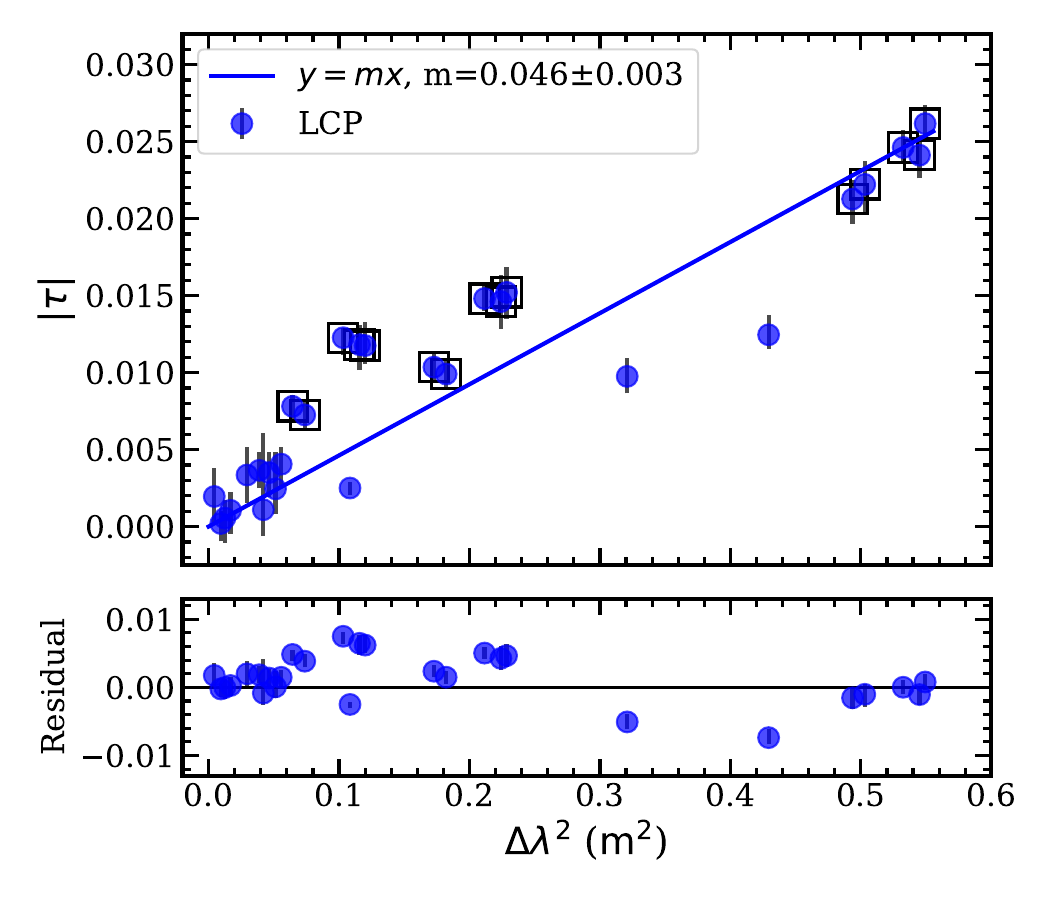}
    \caption{The variation of lag $\tau$ between ECME pulses at two frequencies with the quantity $\Delta\lambda^2=\lambda_1^2-\lambda_2^2$, where $\lambda_1$ and $\lambda_2$ are the wavelengths corresponding to the two frequencies. The RCP pulse considered, (left) was observed near null 1 and the LCP pulse considered (right), was observed near null 2. The linear relation (shown by straight lines) is expected for small frequency ranges and for the case when refraction insider the inner magnetosphere is negligible. The markers surrounded by black squares represent data points that involve measurements using two different instruments (uGMRT and VLA). The residuals show the deviation of the observed $\tau$ vs $\Delta\lambda^2$ relation from linearity.}
    \label{fig:lag_plots}
\end{figure*}

The above exercise provides us with the difference between the pulse arrival phases $\tau$ for each pair of frequencies. The distributions of the lags are shown in Figure \ref{fig:lag_hists}.

We show the lightcurves aligned to the lowest frequency using the measured lag values (with respect to the lowest frequency) in Figure \ref{fig:hd133880_lightcurves_aligned}.
The absolute values of the lags are now plotted against the quantity $\Delta\lambda^2\equiv \lambda_1^2-\lambda_2^2$ ($\lambda$ represents wavelength) in Figure \ref{fig:lag_plots}. 
As mentioned already in \S\ref{sec:intro}, if we approximate the continuous refraction undergone in the stellar magnetosphere by a single refraction at a fixed (in the star's rotating frame of reference) plasma-screen \citep[scenario used by][]{trigilio2011},
the lag values are expected to vary linearly with $\Delta\lambda^2$ over a small range of frequencies \citep[Equation 2 of ][]{das2020a0}. Here we find that the variation of lag with $\Delta\lambda^2$ is clearly non-linear. This is probably not surprising since the linear relation requires near-identical ray paths, which is unlikely to be valid in our case
as we are dealing with a wide range of frequencies (different frequencies are produced at different heights from the stellar surface and traverse different parts of the magnetosphere). Interestingly, we find that the ranges within which the values of lags lie are similar for the RCP and LCP pulses, however, the deviation of the relation between lag and $\Delta\lambda^2$ from linearity is higher for the RCP pulses than that for the LCP pulses. The reduced $\chi^2$ for the straight line model is 9.7 for the LCP data and 52.2 for the RCP data, clearly showing that the straight line model is not valid for either.

In the next section, we attempt to qualitatively understand the significance of the observed difference in the variation of $\tau$ with $\Delta\lambda^2$ for pulses of opposite circular polarizations.

\section{Significance of the non-linear variation of lags with $\Delta\lambda^2=\lambda_1^2-\lambda_2^2$}\label{sec:significance}
Here we will use the 3D framework presented in \citet{das2020a} to simulate the lag between ECME pulses at different frequencies. This framework enables one to obtain ray path for any kind of density distribution in the stellar magnetosphere. It uses the model proposed by \citet{trigilio2011} for the emission of ECME, i.e. it assumes that ECME at any frequency is directed tangential to the relevant auroral ring and parallel to the magnetic equatorial plane. 

Our primary aim is to identify the physical quantities that control the deviation of the variation of the relation between $\tau$ and $\Delta\lambda^2$ from linearity. Since there are several unknowns in the stellar system (e.g. particle acceleration sites, magnetic field topology at the height of radio emission, density distribution inside the inner magnetosphere etc.), we do not attempt to reproduce the observed lags, rather focus on a qualitative understanding via simulations. As shown in the following subsection (\S\ref{subsec:2D_simulation}), this goal can be achieved by considering a magnetosphere with azimuthally symmetric plasma distribution (`2D' simulation). The effect of an azimuthally asymmetric plasma distribution is discussed in \S \ref{subsec:3D_simulation}.

\subsection{Azimuthally symmetric magnetosphere}\label{subsec:2D_simulation}
We consider a star with an axi-symmetric dipolar magnetic field. We take the inclination angle (angle between the stellar rotation axis and the line of sight, $i$) and the obliquity (the angle between the stellar rotation and magnetic dipole axes, $\beta$) of the star to be similar to those of HD\,133880, i.e. $i=65^\circ$ and $\beta=78^\circ$ \citep{bailey2012,kochukhov2017}). Although according to the `Rigidly Rotating Magnetosphere' model \citep[RRM model, ][]{townsend2005}, such high obliquity will imply an azimuthally asymmetric distribution of magnetospheric plasma w.r.t the dipole axis, we will first consider an azimuthally symmetric plasma distribution with an overdense region at the magnetic equator while simulating the lags for the different pair of frequencies. Such a scenario is expected when the rotation and magnetic axes are nearly aligned.
The azimuthally asymmetric case is considered in \S\ref{subsec:3D_simulation}. As will be shown below, the symmetric case allows us to acquire fundamental insights about the spectral dependence of lags inspite of the simplicity of the system.


\citet{das2020b} showed that the magneto-ionic mode of ECME produced by HD\,133880 over the entire frequency range of our observation is extra-ordinary. We assume that the radiation is emitted at the second harmonic. The dipole strength of the star is taken to be 9.6 kG, and the Alfv\'en radius \ra~is taken to be 60 $R_*$\footnote{Note that this estimate has significant uncertainty as this was made assuming a purely dipolar magnetic field, however, \citet{kochukhov2017} reported that the star's magnetic field topology deviates from a dipolar topology. In addition, mass-loss rates of late B-type star are also highly uncertain \citep[e.g. see][for a discussion on different mass-loss rates of B-type stars obtained using different `recipes']{shultz2016}. Also see \citet{petit2013} for a discussion of the consequence of these uncertainties on various magnetospheric characteristics.}
, where $R_*$ represents stellar radius \citep{bailey2012}. We assume that ECME is produced in auroral rings at magnetic field loops which has magnetic equatorial radii lying in the range $60-78\, R_*$. The frequencies for which we simulated ray paths are 400 MHz, 600 MHz, 800 MHz, 1 GHz, 1.2 GHz, 1.4 GHz, 1.6 GHz and 1.8 GHz. 

For this exercise, we consider a density distribution that has a component slowly varying with radial distance, superimposed on it is an overdense plasma disk at the magnetic equator. The disk has a finite width that decreases with radial distance. Also, the density falls steeply away from the equatorial plane. The idea of the overdense plasma disk at the magnetic equator is inspired from the prediction of the RRM model for small obliquities. We obtain such a distribution considering the following analytical function: 
\begin{equation}
    \label{eq:density_function}
  n_p=\frac{n_{p0}}{r}+\frac{n_{p0}}{\sqrt{r}}E(1-\Delta)\exp{\left(-\frac{fz^2}{\sigma^2}\right)},
\end{equation}

\begin{figure}
    \centering
    \includegraphics[width=0.45\textwidth]{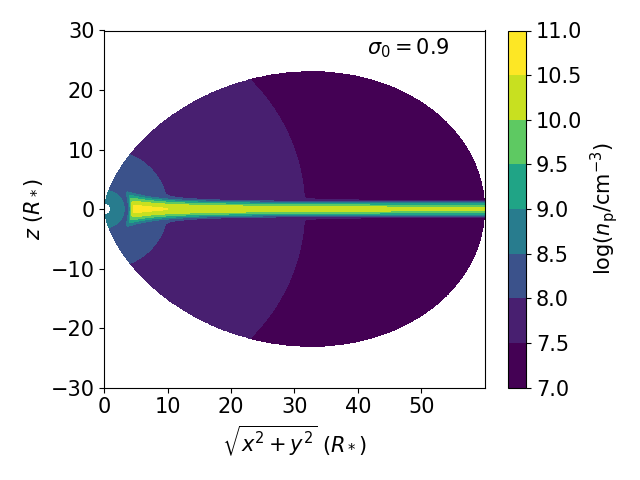}
    \includegraphics[width=0.45\textwidth]{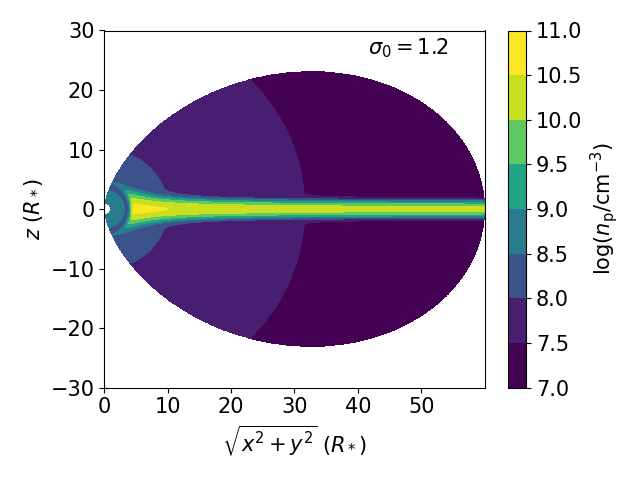}    
    \caption{The density distribution inside the inner magnetosphere represented by a function like that given in Eq. \ref{eq:density_function}. We have used $n_{p0}=10^9\,\mathrm{cm^{-3}}$, $r_0=4.5$ (in the unit of stellar radius), $M=5$, $f=3$, $E=100$ and \ra=60. $\sigma_0=0.9$ and 1.2 for the top and bottom panels respectively.}
    \label{fig:density_IM}
\end{figure}

\begin{figure}
    \centering
    \includegraphics[width=0.4\textwidth]{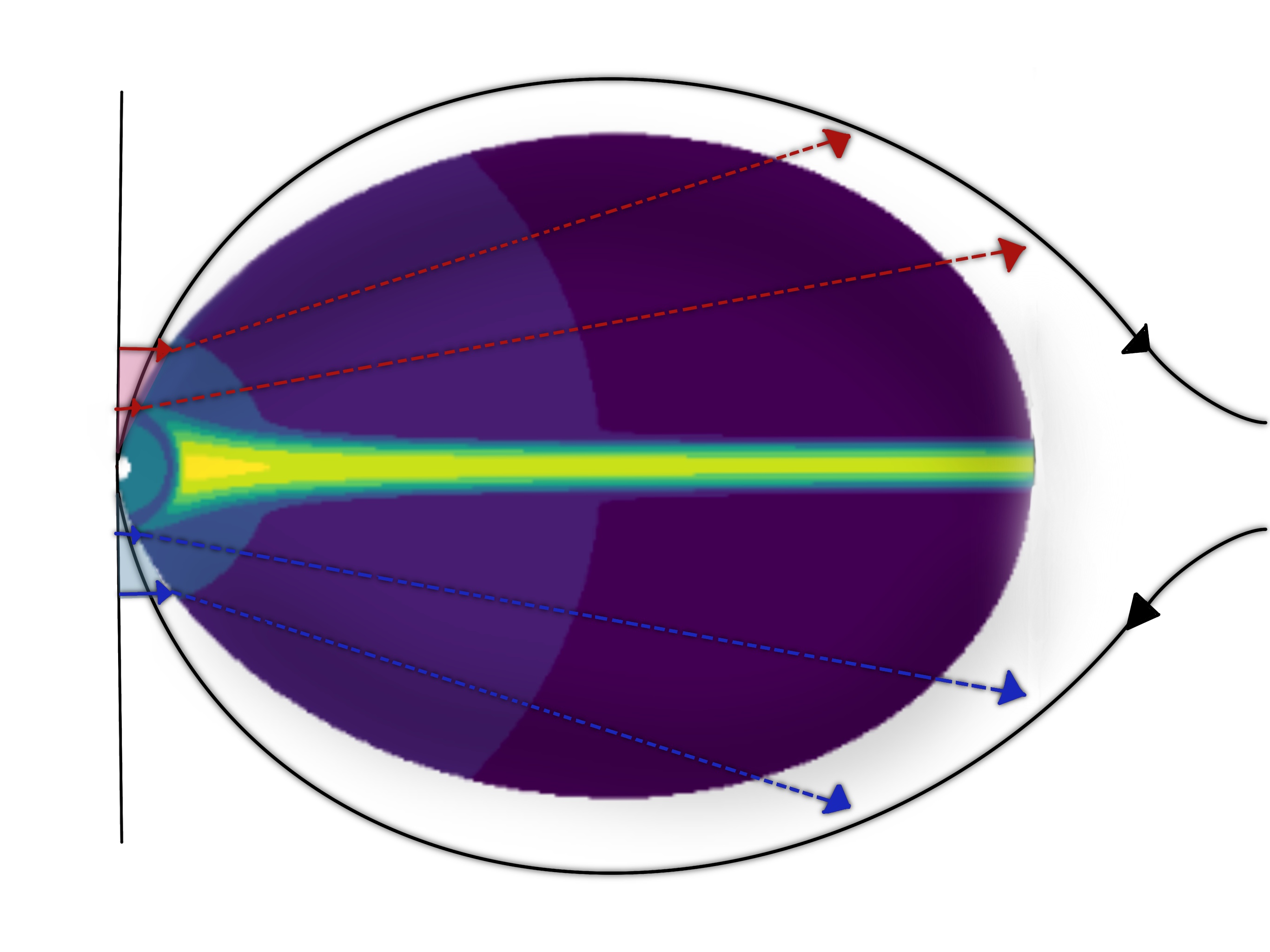}
    \caption{A cartoon diagram illustrating the propagation effects experienced by ECME (shown by arrows) as the radiation passes through the inner magnetosphere (dark shaded region). The inner magnetosphere is surrounded by a region called middle magnetosphere where the magnetic field lines are drawn open near the equator by the stellar wind \citep{trigilio2004}. ECME is produced in auroral rings above the stellar magnetic poles with the height of emission being determined by the emission frequency. The intrinsic direction of emission is assumed to be the same for all frequencies, but the deviations caused by magnetized plasma has a spectral dependence.
    \label{fig:ecme_visualization}
    }
    \label{fig:enter-label}
\end{figure}

Where $n_p$ is the plasma density at a point $(r,\theta)$ (spherical polar coordinates, $z$ axis lies along the magnetic dipole axis) inside the magnetosphere, $n_{p0}$ is a density scaling factor, $E$ controls the magnitude of the overdensity; $\Delta=1/1+\exp(2M(r-r_0))$ is a function that smoothly connects the overdense disk to the background density at point $r=r_0$, where $r_0$ determines the `onset' of the overdensity at the magnetic equatorial plane. Within the exponential function, $f$ is a constant, $\sigma/f$ determines the extent of the density enhancement away from the magnetic equatorial plane. $\sigma$ is taken to be of the form $\sigma_0\exp(R_\mathrm{A}/(x^2+R_\mathrm{A}))$, where $x=r\sin\theta$ and $\sigma_0$ is a constant. Note that all the distances are in units of the stellar radius.
A slice of the inner magnetosphere (along a constant magnetic azimuth) with density distribution given by the above function is shown in Figure \ref{fig:density_IM}. 
An illustration of how the ECME radiations produced over a range of heights above the magnetic poles (corresponding to different frequencies) suffer deviations is shown in Figure \ref{fig:ecme_visualization}.
The lag for a pair of frequencies is calculated using Eq. 7 and 8 of \cite{das2020a}.

\begin{figure*}
    \centering
    \includegraphics[width=0.45\textwidth]{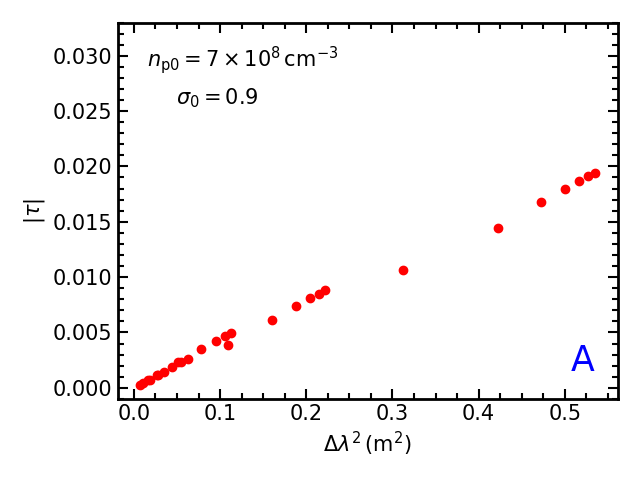}
    \includegraphics[width=0.45\textwidth]{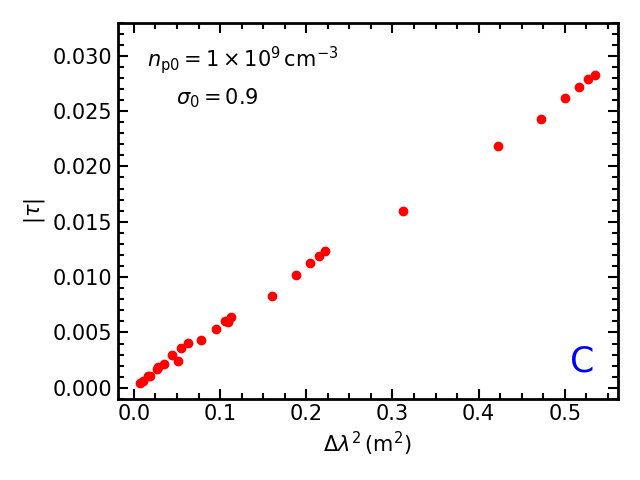}
     \includegraphics[width=0.45\textwidth]{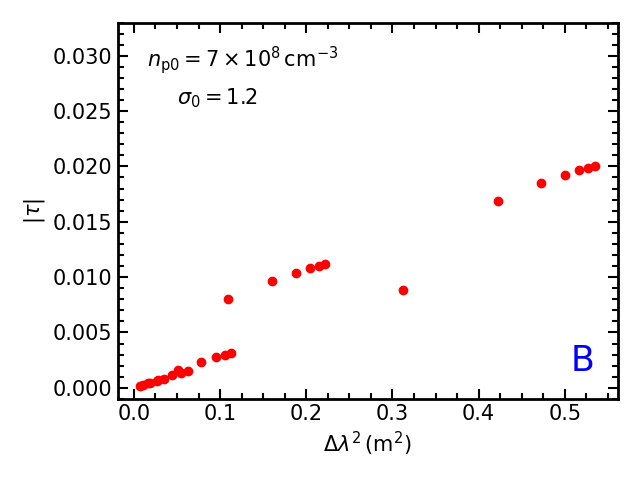}
     \includegraphics[width=0.45\textwidth]{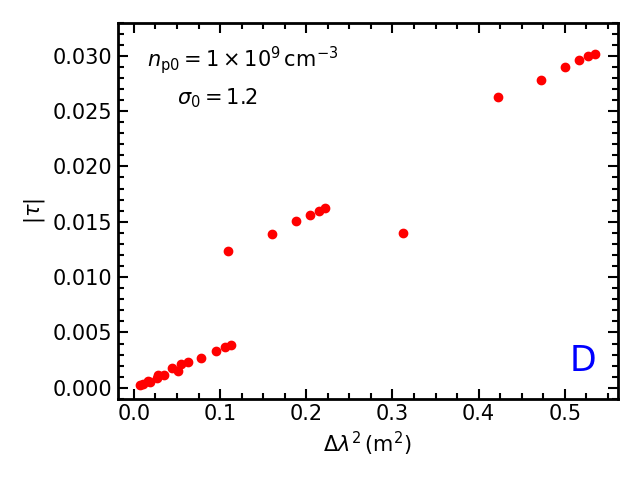}
       \caption{The simulated lags vs $\Delta\lambda^2=\lambda_1^2-\lambda_2^2$ for a star with physical parameters (magnetic field strength, stellar radius and Alfv\'en radius) similar to that of HD\,133880. The density distribution considered has the form given in Eq. \ref{eq:density_function}. In all the plots, we have used $r_0=4.5$, $M=5$, $f=3$, $E=100$ and \ra=60. For the top panels $\sigma_0=0.9$, whereas for the bottom panels $\sigma_0=1.2$. On the other hand, for the left panels, $n_{p0}=7\times10^8\,\mathrm{cm^{-3}}$ and for the right panels, $n_{p0}=1\times10^9\,\mathrm{cm^{-3}}$.}
    \label{fig:sim_lag}
\end{figure*}

We next show the results of our simulation. In Figure \ref{fig:sim_lag}, we show the lags for the different pairs of frequencies as a function of the quantity $\Delta\lambda^2=\lambda_1^2-\lambda_2^2$. For the simulation, we set $r_0=4.5$, $M=5$, $f=3$, $E=100$ and \ra=60. The four plots in Figure \ref{fig:sim_lag} are obtained by considering two values of $n_\mathrm{p0}$
and two values of $\sigma_0$. Our main findings are:
\begin{enumerate}
    \item For a given $\sigma_0$, increasing $n_{p0}$ (i.e. overall increasing the density by a constant factor) primarily increases the values of the lags (e.g. compare panels `A' with `C', or, `B' with `D' of Figure \ref{fig:sim_lag}), without affecting the extent of deviation from linearity.
    \item On the other hand, increasing $\sigma_0$ (keeping $n_{p0}$ fixed) dramatically enhances the extent of deviation from linearity for the relation between lag and $\Delta\lambda^2$, but do not affect the range of the values of lag significantly (e.g. compare panels `A' with `B', or, `C' with `D' of Figure \ref{fig:sim_lag}).
\end{enumerate}

\begin{figure*}
    \centering
    \includegraphics[width=0.45\textwidth]{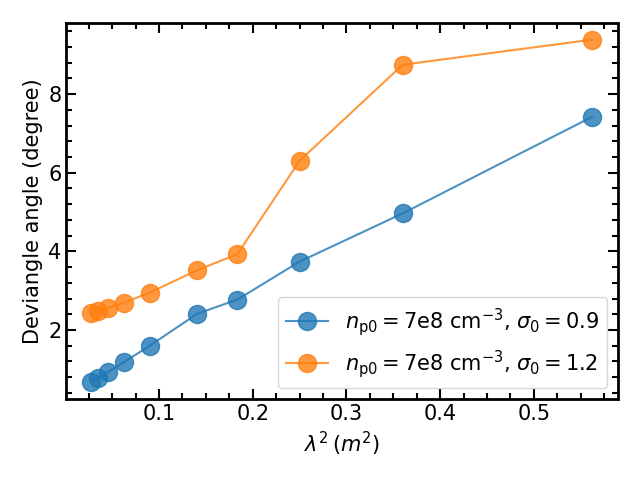}
    \includegraphics[width=0.45\textwidth]{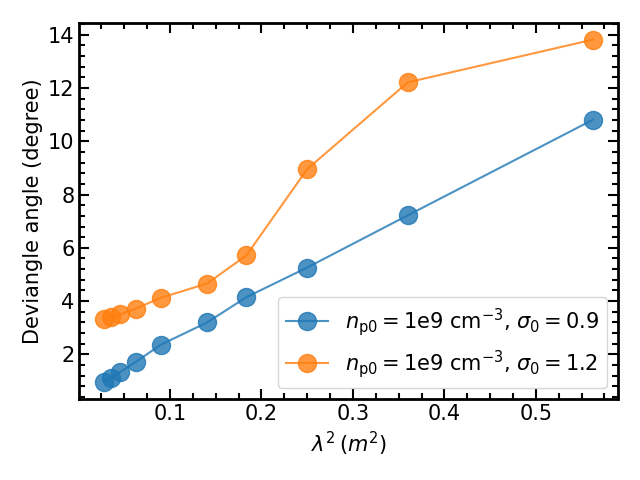}
    \caption{Deviation angles (caused by propagation effects in the stellar magnetosphere) for the simulated ray paths at different frequencies. The magnetospheric density distribution is assumed to have a form given by Eq.\ref{eq:density_function}. The left and right panels correspond to two different values of the absolute density scaling factor $n_{p0}$. For details, refer to \S\ref{subsec:2D_simulation}.}
    \label{fig:dev_nu}
\end{figure*}
For the simulated ray paths, it is possible to obtain the absolute deviation angle caused due to the effect of the magnetospheric plasma. Deviation angle is defined as the angle between the original ray path and that of the final ray path once it emerges out of the stellar magnetosphere. In Figure \ref{fig:dev_nu}, we plot the deviation angles for the different frequencies as a function of square of the wavelengths (we added two new frequencies: 500 MHz and 700 MHz, for a better coverage along $\lambda^2$ axis). For smaller value of $\sigma_0$, the deviation angles increases linearly with $\lambda^2$, however, for the larger value of $\sigma_0$, the relation is no longer linear primarily due to large deviations suffered by the ray paths at frequencies between 400--800 MHz. 

The observed behaviour can be explained by considering the relation among the refractive index, plasma density and frequency of emission. Although in magnetized plasma, refractive index $\mu$ has a complicated functional form \citep[e.g. see][]{lee2013}, to obtain a qualitative understanding, we can consider its simplest form given by $\mu^2=1-(\nu_\mathrm{p}^2/\nu^2)$, where $\nu_\mathrm{p}$ is the plasma frequency that varies as $\sqrt{n_p}$. 
An increase in plasma density leads to a decrease in refractive index and thus an increase in the deviation angle. However, the density required to reduce $\mu$ by a given amount $\Delta \mu$, increases with increasing frequency of emission.
In other words, the higher frequencies are more immune to an increase in plasma density than the lower frequencies. 
At the same time, in case of ECME where the emission frequencies are proportional to the local magnetic field strengths,
lower frequencies are produced farther away from the star. Consequently, for the type of density distributions considered here, they always experience lower plasma densities than that by the higher frequencies. Thus, both the highest and lowest frequencies are less affected due to a change in $n_{p0}$ and $\sigma_0$, and the intermediate frequencies suffer the most. 

When we change $n_\mathrm{p0}$, the densities all over the magnetosphere get scaled up or down, affecting all frequencies in the same way, and thus merely increases or decreases the deviations without affecting the spectral dependence. A change in $\sigma_0$ however, changes the density over a small parts of the magnetosphere, and as the different frequencies are produced at different heights, the amount and type of changes `seen' by different frequencies will also be different in the case. This will then not only change the deviation angles, but also affect their spectral variation as observed in our simulation.

\begin{figure}
    \centering
    \includegraphics[width=0.45\textwidth]{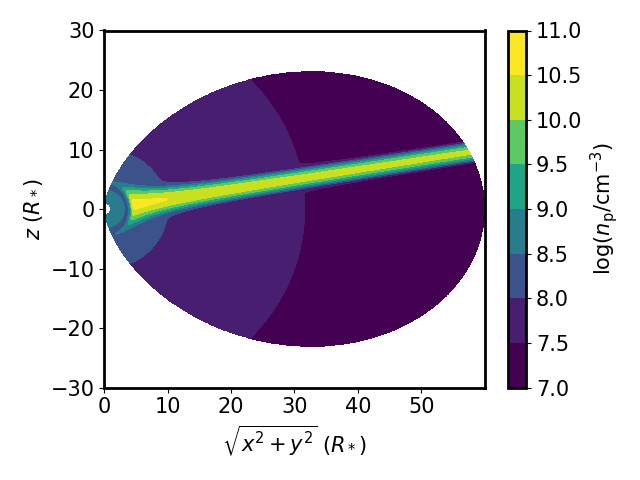}
    \caption{A cartoon diagram illustrating a density distribution inside the inner magnetosphere in which the overdense disk is inclined towards the northern magnetic hemispheres.}
    \label{fig:inclined_density_IM}
\end{figure}

In our observation (Figure \ref{fig:lag_plots}), we find that the range of the lag values for the RCP and the LCP pulses are similar, but the lags for the RCP pulses deviate from the linear relation (with $\Delta\lambda^2$) to a greater extent. 
The above exercise shows that an increased deviation from linearity suggests a thicker disk (larger $\sigma_0$).
As the star HD\,133880 has a large obliquity, it is expected to have a disk like overdensity in its magnetosphere which is not symmetric about the magnetic dipole axis \citep[RRM model of ][]{townsend2005}. Thus, we can infer from our observation and simulation that while both magnetic hemispheres of the star have similar background plasma densities, along the ray paths corresponding to the RCP pulses, the overdense region is more extended as compared to that for the LCP pulses produced in the opposite magnetic hemispheres.
Such a situation can be obtained if the disk is inclined towards the northern magnetic hemisphere for the parts of the magnetosphere through which the radiation corresponding to the observed RCP and LCP pulses passes through. An illustration for this scenario is shown in Figure \ref{fig:inclined_density_IM}.

The above inferences drawn from our `2D' simulation motivates simulation using the plasma distribution predicted by the RRM model.
We present the results for that case in the following subsection.

\subsection{Azimuthally asymmetric magnetosphere}\label{subsec:3D_simulation}
We now consider the case when the stellar magnetosphere has an asymmetric density distribution (function of all three spatial co-ordinates) given by the RRM model. Using a Kepler radius of $2.8\,R_*$ (for HD\,133880) and an obliquity of $78^\circ$, we first construct the density grid following the RRM model \citep{townsend2005}. Note that this grid only provides the density distribution inside the CM. Outside the CM, the density is assumed to be zero. We hence consider a density distribution of the following form:
\begin{align}
    n_\mathrm{p}&=\frac{n_\mathrm{p0}}{r}\left(1+E \Tilde{n}\right)
    \label{eq:3D_density_func}
\end{align}
Where $\Tilde{n}$ is the relative density provided by the RRM model \citep{townsend2005}, $r$ is the radial distance in units of stellar radius, and $n_\mathrm{p0}$ and $E$ have the same significance as that in Eq. \ref{eq:density_function}. The $1/r$ dependence of the number density is motivated from \citet{leto2006}. We fixed $E$ at 100. 

\begin{figure*}
    \centering
    \includegraphics[width=0.45\textwidth]{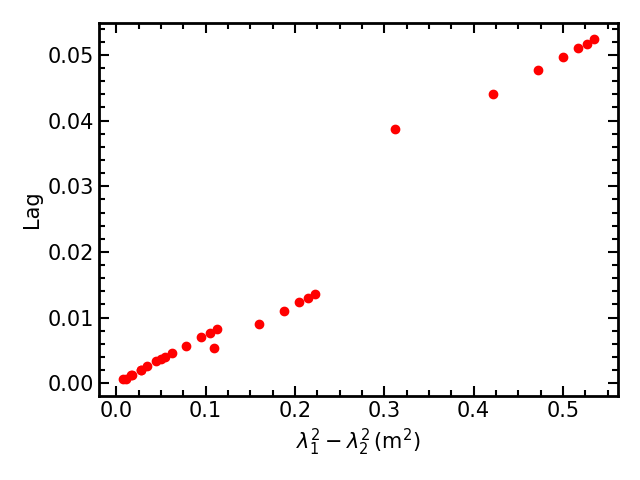}
    \includegraphics[width=0.45\textwidth]{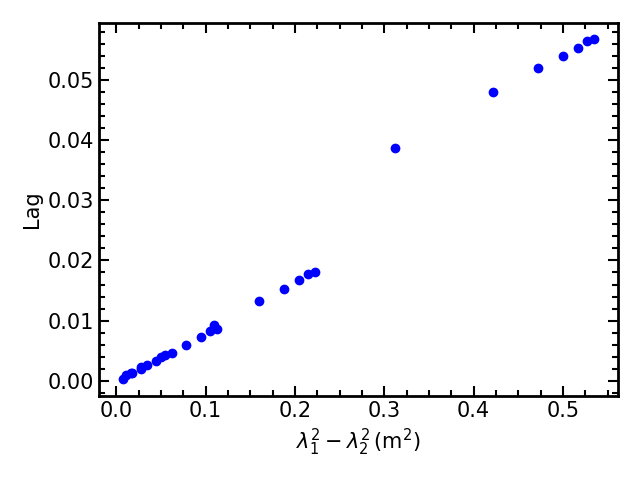}
    \caption{Simulated lags for an azimuthally asymmetric magnetosphere with a density distribution given by the functional form in Eq.\ref{eq:3D_density_func}. We used $n_\mathrm{p0}=7\times 10^8\,\mathrm{cm^{-3}}$ and $E=100$. ECME is assumed to be emitted along field lines with apex radius of $20\,R_*$ at the second harmonic. The polar field strength is taken to be 9.6 kG. The left and right panels show the lags for radiation produced in the northern and southern magnetic hemispheres respectively. Note that we have assumed an inclination angle of $40^\circ$.
    \label{fig:3D_sim_best_case}}
\end{figure*}

By trial and error, we find that a combination of stellar parameters (fixing the polar field strength at 9.6 kG) that allows us to obtain non-identical lags for the two magnetic hemispheres (with values similar to the observed ones) is when the ECME is produced along field lines with equatorial radius of $20\,R_*$ at the second harmonic, with $n_\mathrm{p0}=7\times 10^8\,\mathrm{cm^{-3}}$ and an inclination angle of $40^\circ$ (changing the inclination angle of the rotation axis is equivalent to tweaking the density distribution).
The corresponding lags are shown in Figure \ref{fig:3D_sim_best_case}.

\begin{figure*}
    \centering
    \includegraphics[width=0.45\textwidth]{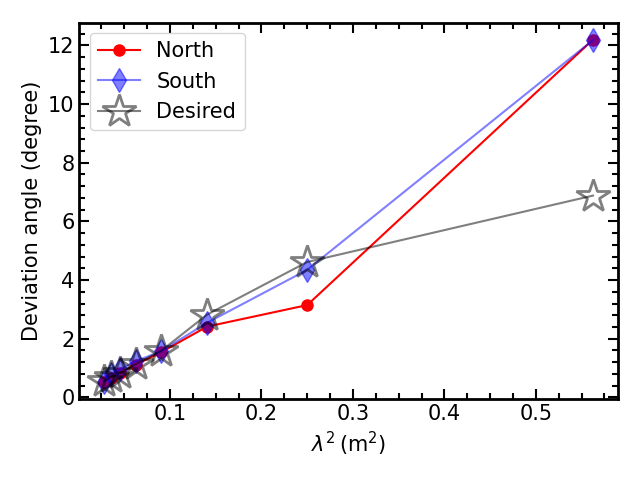}
    \includegraphics[width=0.45\textwidth]{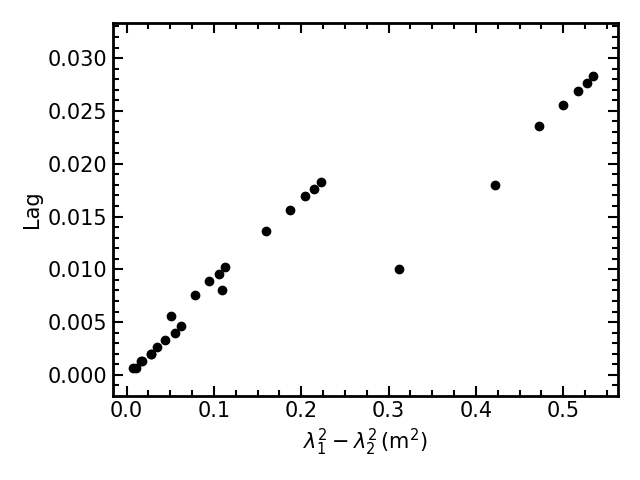}
    \caption{\textit{Left:} Deviation angles (red circles and blue diamonds) for the ray paths corresponding to Figure \ref{fig:3D_sim_best_case}. The markers in grey show the desired deviation angle to obtain a lag variation shown on the right panel, which is similar to what we observed (left panel of Figure \ref{fig:lag_plots}).
    \label{fig:3D_devang_desired}
    }
\end{figure*}

While the simulated lags match our observation in the sense that they do not vary linearly with $\Delta\lambda^2$ and their extent of deviation are different for the two magnetic hemispheres, the type of variation exhibited by the simulated lags is in stark contrast to observation. For the observed pulses from the Northern magnetic hemisphere,
the lags at smaller $\Delta\lambda^2$ over-predict the lags at larger $\Delta\lambda^2$, whereas for the simulated lags, the lags at smaller $\Delta\lambda^2$ under-predict the lags at higher $\Delta\lambda^2$.
(e.g. see the left panels of Figures \ref{fig:lag_plots} and \ref{fig:3D_sim_best_case}). The reason behind the difference can be understood by plotting the corresponding deviation angles (left panel of Figure \ref{fig:3D_devang_desired}). From \S\ref{subsec:2D_simulation}, we learned that to reproduce the observed lag variation, we require the intermediate frequencies to get affected the most because of enhanced density. On the contrary, for the RRM model, we find that the lowest frequency is affected the most by the overdense disk. We test this idea by manually reducing the deviation suffered by the lowest frequency (400 MHz) and increasing the deviation suffered by the intermediate frequencies (600 and 800 MHz, shown by black stars in the left panel of Figure \ref{fig:3D_devang_desired}). As shown in the right panel, the new relation between lag and $\Delta\lambda^2$ is very similar to what we observed for the RCP pulses (left panel of Figure \ref{fig:lag_plots}).

Thus, the main takeaway from our simulation is that strong deviation of the relation between lags and $\Delta\lambda^2$ from linearity is a tell-tale signature of encountering a localized overdense region in the stellar magnetosphere.
The observed variation (especially for the RCP pulses) suggests a density structure that affects the frequencies between 400--800 MHz the most. Such a case is possible if the density along the disk is more centrally concentrated than that predicted by the RRM model. Indeed, recent MHD simulation of oblique stellar magnetosphere has found that the density decreases as $1/r^5$ rather than the RRM scaling that goes as $1/r^3$ \citep{ud-doula2023}. In addition, our model favours the scenario when the magnetic field lines producing ECME lie much closer to the star than the estimated Alfv\'en radius of the star. This is again consistent with the recent finding that sites of particle acceleration are located inside the Alfv\'en radius \citep{leto2021,shultz2022}.

\section{Discussion and summary}\label{sec:summary}
In this paper, we demonstrate via the example of HD\,133880 that wideband observations of coherent radio emission from hot magnetic stars allow us to probe the three dimensional density structures in the stellar magnetosphere. 
The key quantity is the lag $\tau$ between a pair of frequencies and its variation with $\Delta\lambda^2$. 
We present a strategy to robustly estimate the lags and the associated error bars from the observed lightcurves. The variation of $\tau$ with $\Delta\lambda^2$ is found to strongly deviate from linearity, especially for the RCP pulses. By performing simulations of ray paths for stellar magnetospheres with density distributions similar to that predicted by the RRM model, we interpret this deviation as a consequence of propagating through an overdense plasma disk that affects the intermediate frequencies the most.

This is a first step towards making coherent radio emission a mainstream magnetospheric probe. Here we focus on the qualitative reproduction of the observed spectral variation of the differences in pulse arrival times, and in that process, develop an understanding of what causes the observed variation. In the future, we aim to refine this procedure so as to be able to make a quantitative comparison between observations and simulations.

The technique described here makes the crucial assumption that the observed lags are attributed to propagation effects in the magnetosphere.
It is hence highly desirable that the wideband observations are acquired simultaneously or near-simultaneously, so that the measured quantities are not affected by any temporal variability in the ECME phenomenon. For the data used in this paper, the data were acquired over a period of six months (May -- November 2019), and based on the available data on this star from different epochs, we estimate the rotational phase offsets caused by non-simultaneous data to be insignificant for the result presented in this paper (\S\ref{sec:non-simultaneity}).
But the most ideal way to ensure that the measure lags are intrinsic is to conduct simultaneous ultra-wideband observations unless the rotational period evolution of the star is characterized.

The requirement for simultaneous acquisition of wideband data becomes much more stringent for modelling the ECME spectra in order to extract density information. In principle, the extent of free-free absorption in the spectrum can also provide information regarding magnetospheric plasma densities. 
This strategy has been successfully used for radio-sources that emit via incoherent, broadband emission mechanism such as synchrotron \citep[e.g.][etc.]{tingay2003,callingham2015,nayana2018}.
For coherent mechanism such as ECME, the key problem, however, is that the intrinsic spectral shape of the phenomenon is not known. Also, since the emission sites are different for the different frequencies, and the fact that coherent radio emission is extremely sensitive to the local conditions, it is not yet clear whether the spectral shape over a wide frequency range (such as the one used in this work) is time-invariant. Finally, the pulses are known to exhibit variable peak flux densities (e.g. see Figure \ref{fig:old_new_lc}), so that a spectrum obtained by combining multi-frequency observations acquired at different epochs is not necessarily meaningful. The use of temporal information, thus offers the most promising strategy to extract magnetospheric density information.

Though relatively robust to the effect of non-simultaneous data, the method described here have additional limitations that need to be addressed in the future. This include
the assumption of an axi-symmetric dipolar magnetic field, and ignoring the finite bandwidths of the observed flux densities. HD\,133880 is known to have unequal polar field strengths \citep{kochukhov2017}. In the future, we plan to relax the assumption of dipolar magnetic field in the 3D framework of \citet{das2020a}, and also investigate the effect of different bandwidths for different central frequencies. 

Despite the several limitations that exist at the current stage, it is encouraging to find that the qualitative inferences (non-azimuthally symmetric plasma distribution, centrally concentrated density, and non-importance of $R_\mathrm{A}$) are consistent with recent ideas proposed for centrifugal magnetospheres. But the uniqueness of ECME lies in its sensitivity to 
changes in small spatial scales (equivalently, its sensitivity to the `details'), attributed to its high directivity and intrinsically narrow bandwidth. The four pulses of ECME, sampled over a wide range of frequencies constitute a treasure-trove of information allowing us to compare the `top', `bottom', `left' and `right' sides of the magnetosphere (see Figure \ref{fig:ecme_ideal}). 
To fully exploit this unique probe, it will be important to investigate spectral variation of lags for all four pulses for a sample of MRPs to quantify the deviation from the prediction of the RRM model. Such wideband observations will also be able to provide greater insights as to the validity of the assumption that the lags can be wholly attributed to propagation effects. 
At the moment, similar wideband observations have been reported for three more MRPs: CU\,Vir \citep{das2021}, HD\,35298 \citep{das2022c} and HD\,142990 \citep{das2023}. However, for both CU\,Vir and HD\,142990, the pulse profiles are found to vary widely with frequencies. For the latter, the possible reason is speculated to be the fact that the stellar rotation and magnetic dipole axes are misaligned by nearly $90^\circ$. CU\,Vir, however, is more similar to HD\,133880, and the reason behind the observed frequency dependence of ECME is not yet clear. In case of HD\,35298, the overall behaviour of the pulses with frequencies is closer to the ideal picture, but the main challenge is that its pulses are significantly fainter (owing to its greater distance), making it difficult to extract lightcurves at high frequency and time resolutions.

The phenomenon of frequency dependent pulse-profile, though much less prominent, is also observed for HD\,133880. From Figure \ref{fig:hd133880_lightcurves}, it is clear that at lower frequencies, the rotational phase corresponding to the maximum flux density is significantly offset from that of the pulse-centre (e.g. compare the RCP pulse at 593 MHz with that at 1808 MHz). The significance of this phenomenon is unclear at this stage, and could be due to a combination of both intrinsic and extrinsic effects. Potential reasons include intrinsically frequency dependent beaming pattern, manifestations of non-dipolar magnetic field topology, or frequency dependent absorption (e.g. free free absorption) along different lines of sight. Obtaining similar observations for a larger sample of stars is likely to be useful to provide greater insights into this phenomenon.

For stars that produce both H$\alpha$ and ECME (e.g. HD\,142990), the use of both will provide stringent constraints on the density distribution, and is also likely to improve understanding of the phenomena themselves. These objects will be especially important to `calibrate' ECME as a probe, so that it can be applied for cooler magnetic early-type stars on its own.


We conclude by noting that in order to achieve the ultimate goal of acquiring quantitative information of the host star from ECME, it is of utmost importance to expand the sample of MRPs to span a broader stellar parameter phase-space and also focus on wideband characterization of the phenomenon. 


\begin{acknowledgments}
We thank the referee for their constructive comments that helped us to improve the manuscript.
BD thanks Surajit Mondal and Apurba Bera for useful discussion. 
We thank the staff of the GMRT and the National Radio Astronomy Observatory (NRAO) that made our observations
possible. The GMRT is run by the National Centre for Radio Astrophysics of the Tata Institute of Fundamental Research. The National Radio Astronomy Observatory is a facility of the National Science Foundation operated under cooperative agreement by Associated Universities, Inc. This research has made use of NASA's Astrophysics Data System.
\end{acknowledgments}

%

\vspace{5mm}
\facilities{uGMRT, VLA}


\software{astropy \citep{2013A&A...558A..33A,2018AJ....156..123A},  
          numpy \citep{harris2020array}, 
          scipy \citep{2020SciPy-NMeth},
          CASA \citep{mcmullin2007}
          }



\appendix
\restartappendixnumbering 
\section{Effect of non-simultaneous data}\label{sec:non-simultaneity}
For reliable estimates of the lags between different frequencies, it is desirable to acquire the data over the entire frequency range under consideration simultaneously, unless it is well-established that the rotational phases of arrival of the pulses are time-invariant. This is, however, often not practical. In this section, we attempt to estimate an uncertainty in the lag estimates owing to the fact that the data used were not simultaneous.


The most common reason invoked for evolution of the pulse-arrival phases is variable stellar rotation period. The first discovered MRP CU\,Vir is known to exhibit variable rotation period, and its radio pulses have been observed to arrive at seemingly different rotational phases over timescales of a year \citep{trigilio2008}. In addition, \citet{ravi2010} also suggested that the emission sites may themselves drift in time causing the pulses to appear at slightly different rotational phases when observed at different epochs.

In case of our target HD\,133880, \citet{das2020b} reported that there is a phase offset of $\approx$ 0.02 cycles between the band 4 pulses observed in 2019 around Null 2 and the corresponding pulses observed in 2016 \citep{das2018}. In view of this, we choose the data that are most closely spaced in time so as to minimize the error incurred due to offset caused by effects other than propagation effects.

\begin{figure}
    \centering
    \includegraphics[width=0.85\textwidth]{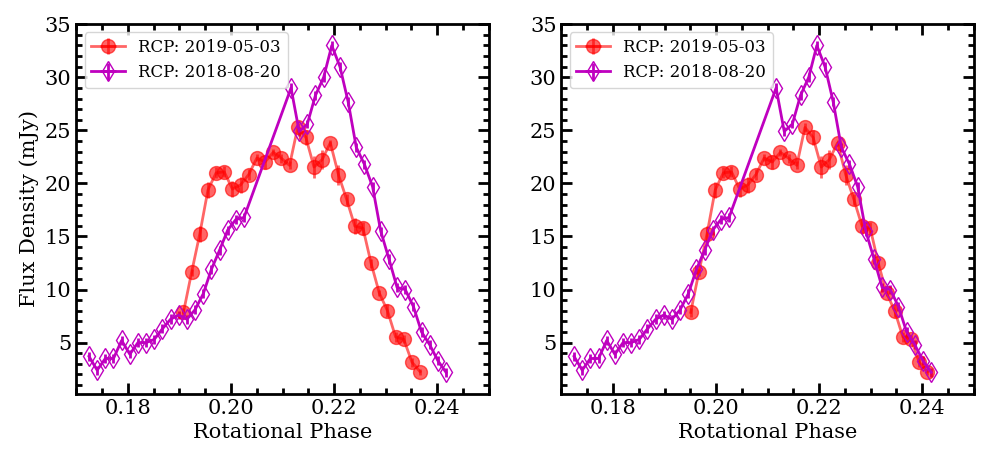}
    \caption{\textit{Left:} The RCP pulse near Null 1 at 687 MHz, observed at two epochs. The data are phased assuming a constant rotation period of 0.877483 days and a reference epoch of $\mathrm{HJD_0}=2445472.0$ days. \textit{Right:} The phases of the lightcurve from 2019 are increased according to the estimated offset between the original lightcurves (see \S\ref{sec:non-simultaneity}).}
    \label{fig:old_new_lc}
\end{figure}

The VLA data over 1--4 GHz are simultaneous for each magnetic null. The uGMRT data near Null 1 (RCP) are separated by less than half a month, and can be treated as near-simultaneous.
However, the uGMRT data around Null 2 (LCP) are separated by 4.5 months, and the uGMRT and VLA data are separated by nearly 4 months (Null 1) and 7.5 months (Null 2). In order to estimate the shifts in the pulse arrival phases because of stellar rotation period evolution, the RCP pulse observed around Null 1 on 2019--5--03 (this work) with that observed on 2018--08--20 \citep{das2020b} are compared (left of Figure \ref{fig:old_new_lc}). The pulses are clearly offset. To estimate the offset, we employ the same procedure as the one applied to calculate the lags between different frequencies (\S\ref{sec:finding_lags}). The `offset' comes out to be 0.0043 with a 90\% confidence interval of 0.0042--0.0046. The aligned lightcurves according to this estimated offset are shown in the right of Figure \ref{fig:old_new_lc}.

The phase-offset of 0.0043 cycles is observed over 256 days that gives $0.0043/256=1.7\times 10^{-5}$ cycles of phase-offset between two observations separated by a day. Interestingly, if we consider the phase-offset reported by \citet{das2020b}, the phase-offset per day comes out to be $1.6\times 10^{-5}$ cycles per day. Using this number, the phase-offset between observation at two epochs separated by $n$ days is calculated as $n\times 1.7\times 10^{-5}$ cycles. In Table \ref{tab:estimated_lags}, we show the estimated lags $\tau_0$, their 90\% confidence intervals $\tau_{90}$ and the offset calculated in this way. Note that this offset is not a random uncertainty in the sense that a longer gap between two epochs of observation will always lead to an increase in the observed offset (for $\nu_2>\nu_1$) for the RCP and LCP pulses near Null 1 and Null 2 respectively.

In all cases, we find that the $\tau_0$ is significantly larger than the offset.
`Correcting' the lags using the offset does not affect the non-linear variation with $\Delta\lambda^2$.

\begin{deluxetable*}{cccccccccccc}
\tabletypesize{\small}
\centering
\tablecaption{The estimated lags ($\tau_0$) and the associated uncertainties for the RCP (Null 1) and LCP pulse (Null 2) considered in this work. The column labelled $\tau_{90}$ lists the 90\% confidenece intervals. The separation column shows the separation in days between the observations at the two frequencies $\nu_1$ and $\nu_2$. The column labelled `Offset' is the estimated phase-offset incurred by the change in stellar rotation period. Note that the lowest frequency for the RCP pulses (Null 1) is 397.5 MHz and that for the LCP pulse (Null 2) 395 MHz. In the table, we list the average of the two values. See \S\ref{sec:non-simultaneity} for details.\label{tab:estimated_lags}}  
\tablehead{
$\nu_1$ & $\nu_2$& \multicolumn{4}{c}{Null 1 (RCP)}& \multicolumn{4}{c}{Null 2 (LCP)}\\
 &  & Separation & $\tau_0$ & $\tau_{90}$& Offset & Separation & $\tau_0$ & $\tau_{90}$& Offset\\
(MHz) & (MHz) & (days) & (cycles)& (cycles) & (cycles) & (days) & (cycles)& (cycles) & (cycles)
}
\startdata
\hline
396 & 593 & 14 & 0.0049 & (0.0042--0.0052) & 0.0002 & 138 & 0.0098  &(0.0086--0.0110) & 0.0023\\
396 & 781 & 14 & 0.0109 & (0.0102--0.0114) & 0.0002 & 138 & 0.0124  &(0.0115--0.0137) & 0.0023 \\
396 & 1040 & 113 & 0.0191 & (0.0176--0.0201) & 0.0019 & 229 & 0.0213 & (0.0196--0.0225) & 0.0039\\
396 & 1104 & 113 & 0.0194 & (0.0148--0.0202) & 0.0019 & 229 & 0.0222  &(0.0204--0.0237) & 0.0039\\ 
396 & 1424 & 113 & 0.0240 & (0.0230--0.0248) & 0.0019 & 229 & 0.0246 & (0.0236--0.0257) & 0.0039\\
396 & 1680 & 113 & 0.0252 & (0.0245--0.0266) & 0.0019 & 229 & 0.0241 & (0.0226--0.0254) & 0.0039\\
396 & 1808 & 113 & 0.0251 & (0.0216--0.0264) & 0.0019 & 229 & 0.0262 & (0.0250--0.0274) & 0.0039\\
593 & 781 & 0 & 0.0067 & (0.0063--0.0070) & 0 & 0 & 0.0025  &(0.0020--0.0029) & 0\\
593 & 1040 & 127 & 0.0162  &(0.0156--0.0164) & 0.0022 &  91 & 0.0104  &(0.0094--0.0113) & 0.0015\\
593 & 1104 & 127 & 0.0166  &(0.0158--0.0172) & 0.0022 & 91 & 0.0099 & (0.0089--0.0109) & 0.0015\\
593 & 1424 & 127 & 0.0202  &(0.0196--0.0206) & 0.0022 & 91 & 0.0148 & (0.0140--0.0157) & 0.0015\\
593 & 1680 & 127 & 0.0207  &(0.0200--0.0216) & 0.0022 & 91 & 0.0146 & (0.0128--0.0164) & 0.0015\\
593 & 1808 & 127& 0.0219  &(0.0205--0.0229) & 0.0022 & 91 & 0.0152  &(0.0135--0.0169) & 0.0015\\
781 & 1040 & 127 & 0.0081  &(0.0075--0.0088) & 0.0022 & 91 & 0.0078 & (0.0068--0.0085) & 0.0015\\
781 & 1104 & 127 & 0.0078  &(0.0064--0.0092) & 0.0022 & 91 & 0.0072 & (0.0063--0.0083) & 0.0015\\
781 & 1424 & 127 & 0.0130 & (0.0120--0.0137) & 0.0022 & 91 & 0.0123  &(0.0111--0.0129) & 0.0015\\
781 & 1680 & 127 & 0.0144  &(0.0135--0.0153) & 0.0022 & 91 & 0.0118  &(0.0102--0.0131) & 0.0015\\
781 & 1808 & 127 & 0.0139  &(0.0109--0.0150) & 0.0022 & 91 & 0.0118  &(0.0105--0.0133) & 0.0015\\
1040 & 1104 & 0 & 0.0006  &(0.0003--0.0012) & 0 & 0 & 0.0002  &(-0.0010--0.0010) & 0\\
1040 & 1424 & 0 & 0.0039 & (0.0035--0.0044) & 0 & 0 & 0.0036 &(0.0025--0.0049) & 0 \\
1040 & 1680 & 0 & 0.0050 & (0.0034--0.0060) & 0 & 0 & 0.0024 &(0.0008--0.0048) & 0 \\
1040 & 1808 & 0 & 0.0070  &(0.0058--0.0078) & 0 & 0 & 0.0040 &(0.0026--0.0052) & 0 \\
1104 & 1424 & 0 & 0.0030  &(0.0026--0.0038) & 0 & 0 & 0.0033 &(0.0015--0.0052) & 0 \\
1104 & 1680 & 0 & 0.0042 & (0.0027--0.0063) & 0 & 0 & 0.0011 &(-0.0006--0.0061) & 0 \\
1104 & 1808 & 0 & 0.0064  &(0.0051--0.0075) & 0 & 0 & 0.0035 &(0.0019--0.0049) & 0 \\
1424 & 1680 & 0 & 0.0000  &(-0.0001--0.0016) & 0 & 0 & -0.0005& (-0.0022--0.0007) & 0\\ 
1424 & 1808 & 0 & 0.0025  &(0.0012--0.0034) & 0 & 0 & 0.0010 &(-0.0005--0.0023) & 0 \\
1680 & 1808 & 0 & 0.0000  &(-0.0024--0.0024) & 0  & 0 & 0.0020& (0.0001--0.0038) & 0 \\
\enddata
\end{deluxetable*}

\section{Histograms of the lag distributions}\label{sec:lag_histogram}
\begin{figure*}
    \includegraphics[width=0.23\textwidth]{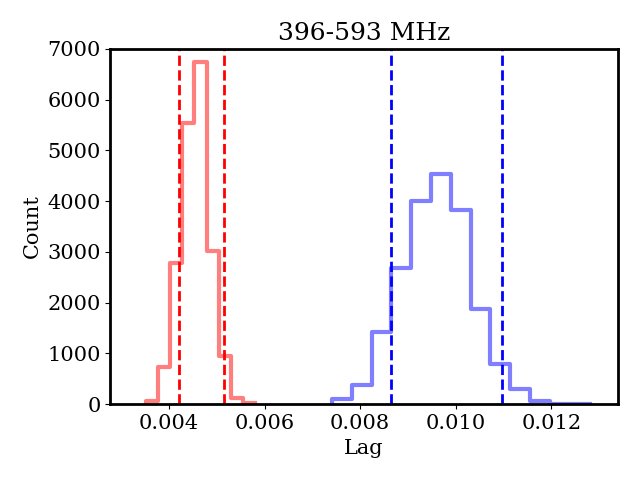}
    \includegraphics[width=0.23\textwidth]{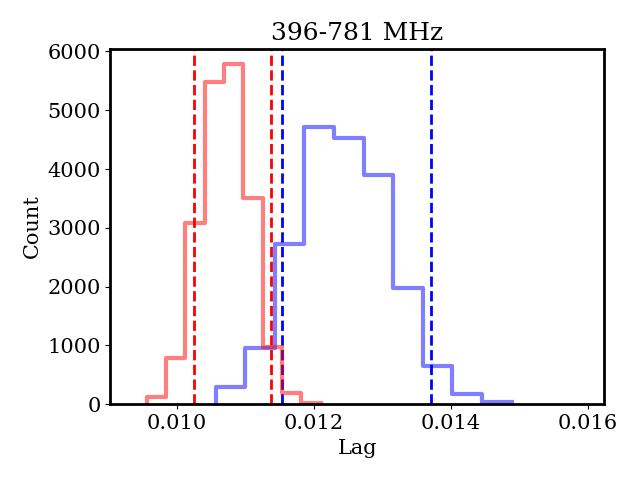}
    \includegraphics[width=0.23\textwidth]{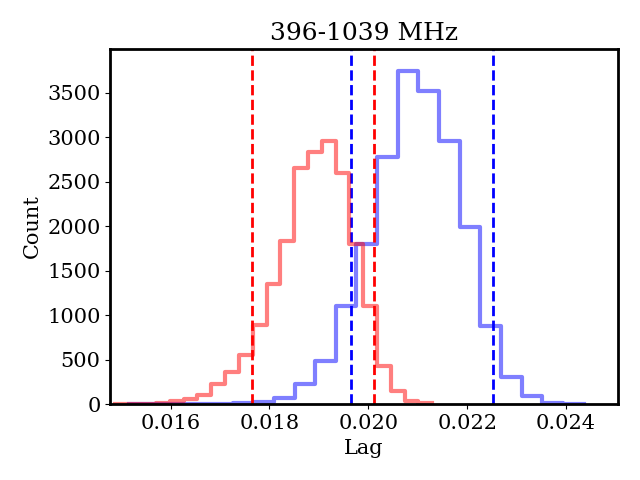}
    \includegraphics[width=0.23\textwidth]{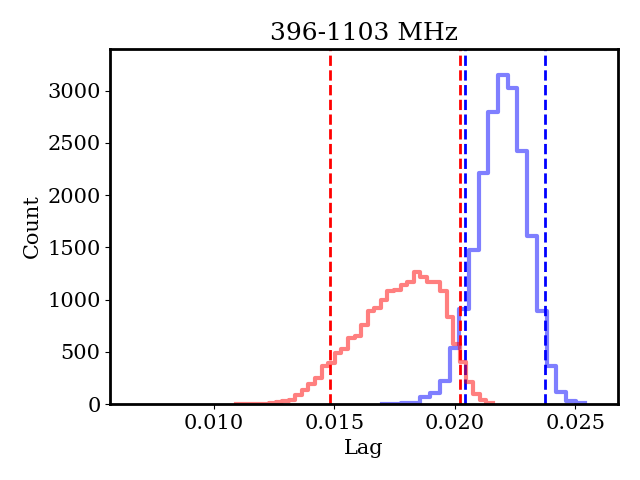}
    \includegraphics[width=0.23\textwidth]{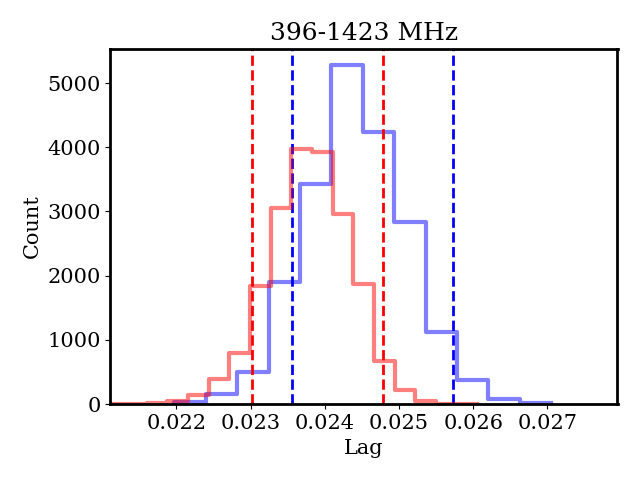}
    \includegraphics[width=0.23\textwidth]{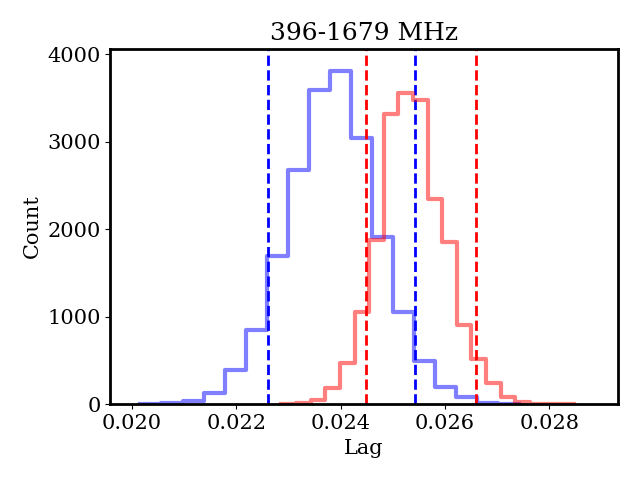}
    \includegraphics[width=0.23\textwidth]{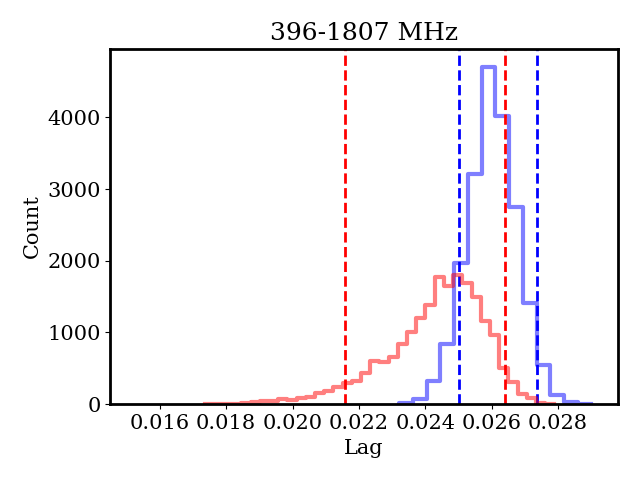}
    \includegraphics[width=0.23\textwidth]{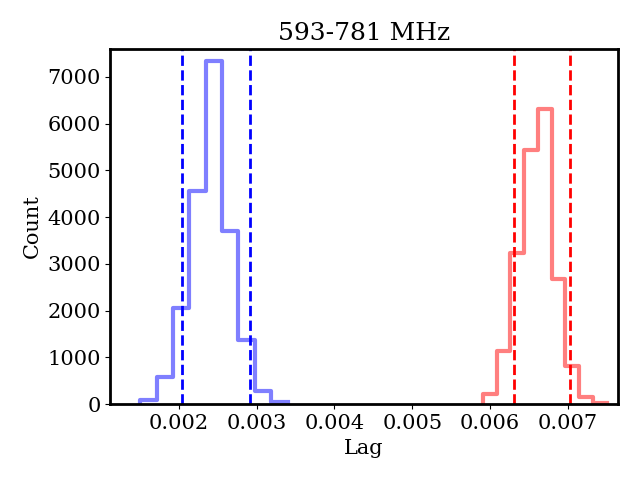}
    \includegraphics[width=0.23\textwidth]{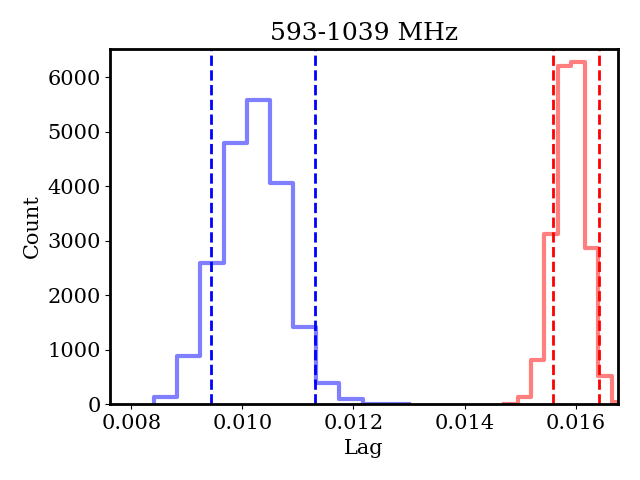}
    \includegraphics[width=0.23\textwidth]{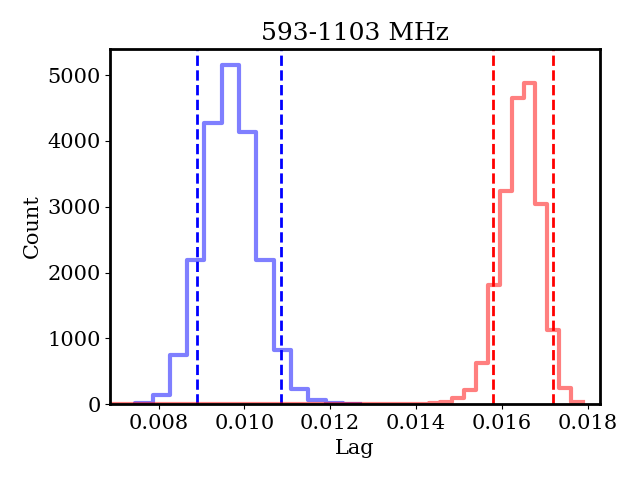}
    \includegraphics[width=0.23\textwidth]{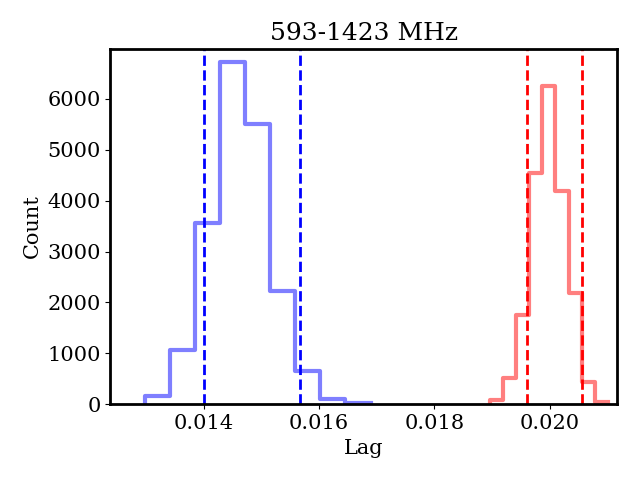}
    \includegraphics[width=0.23\textwidth]{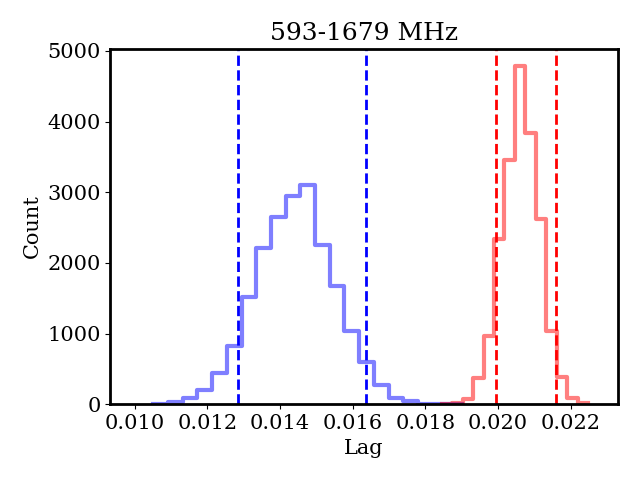}
    \includegraphics[width=0.23\textwidth]{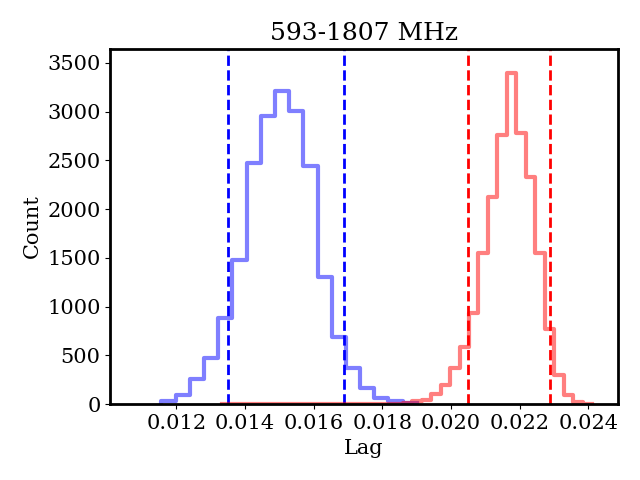}
    \includegraphics[width=0.23\textwidth]{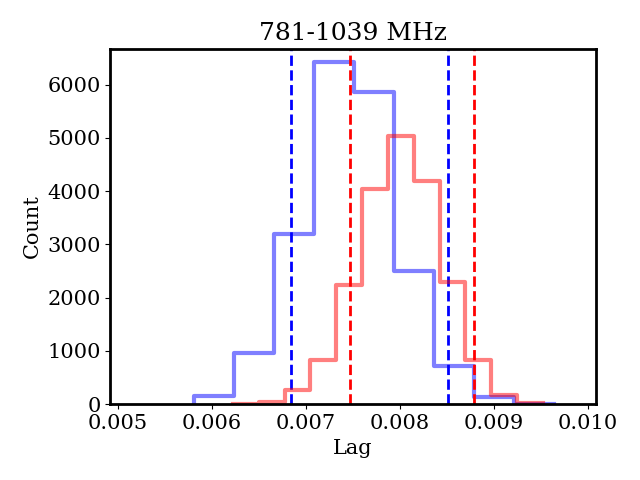}
    \includegraphics[width=0.23\textwidth]{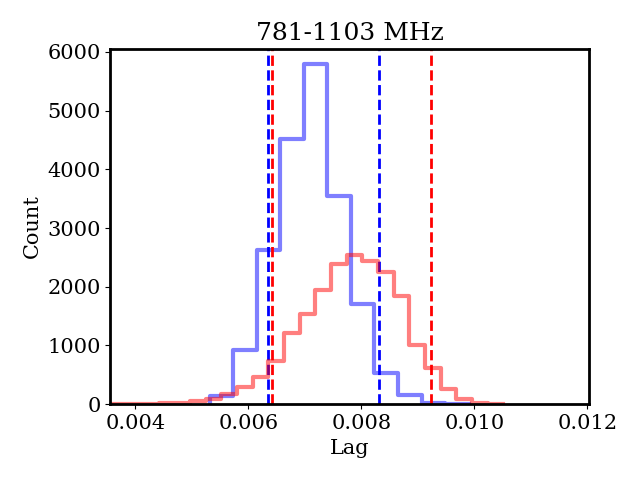}
    \includegraphics[width=0.23\textwidth]{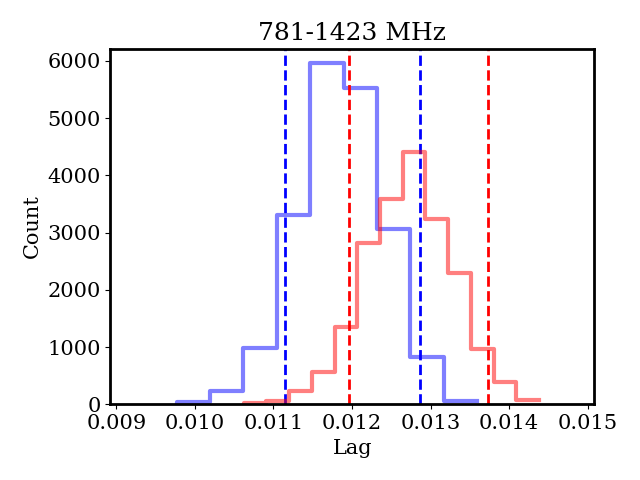}
    \includegraphics[width=0.23\textwidth]{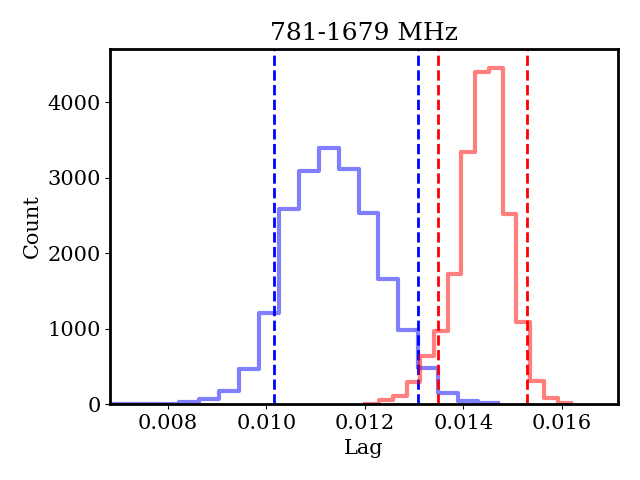}
    \includegraphics[width=0.23\textwidth]{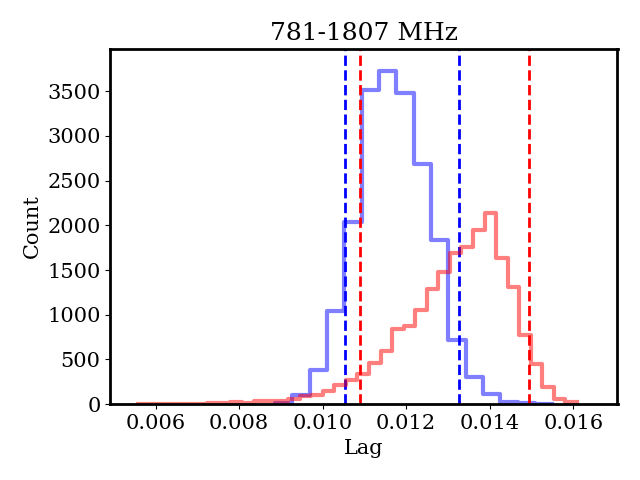}
    \includegraphics[width=0.23\textwidth]{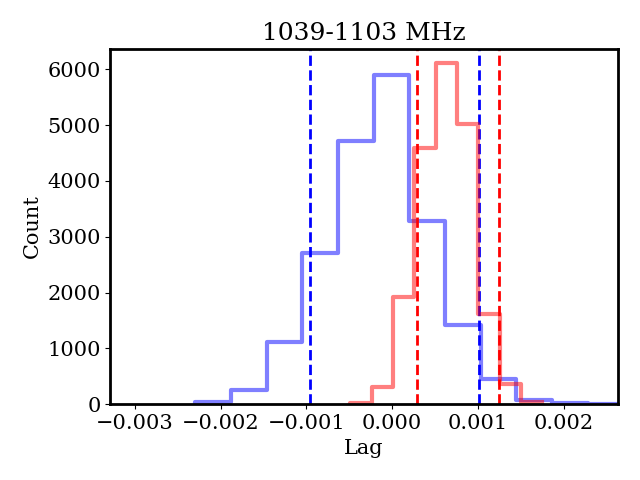}
    \includegraphics[width=0.23\textwidth]{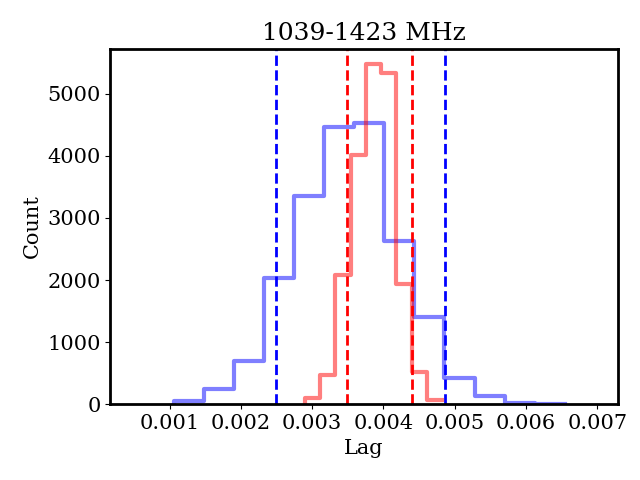}
    \includegraphics[width=0.23\textwidth]{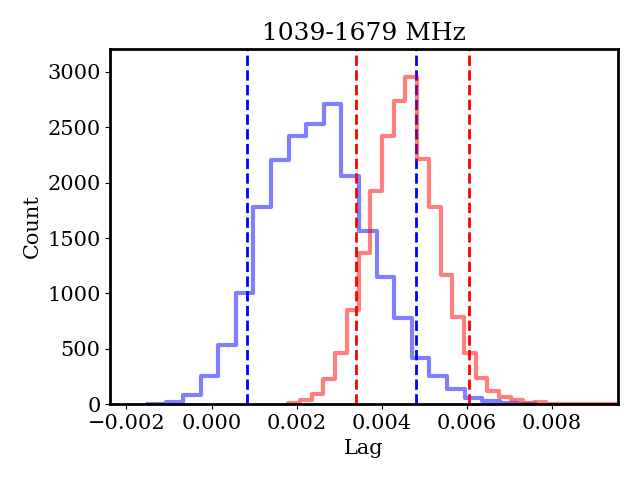}
    \includegraphics[width=0.23\textwidth]{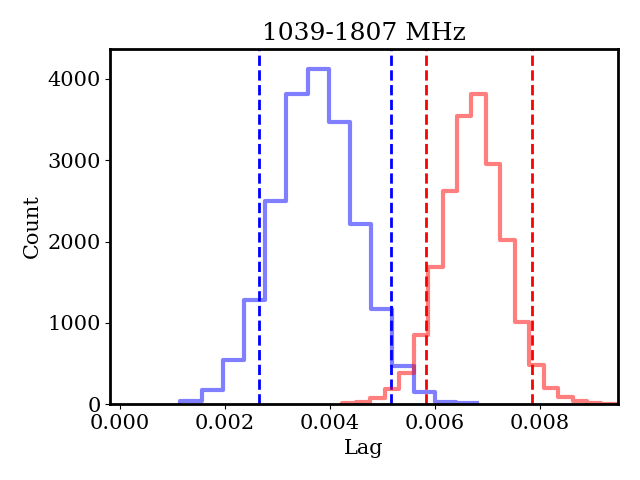}
    \includegraphics[width=0.23\textwidth]{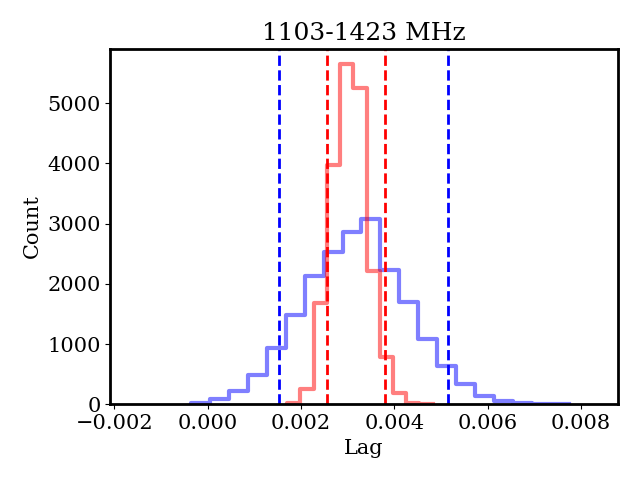}
    \includegraphics[width=0.23\textwidth]{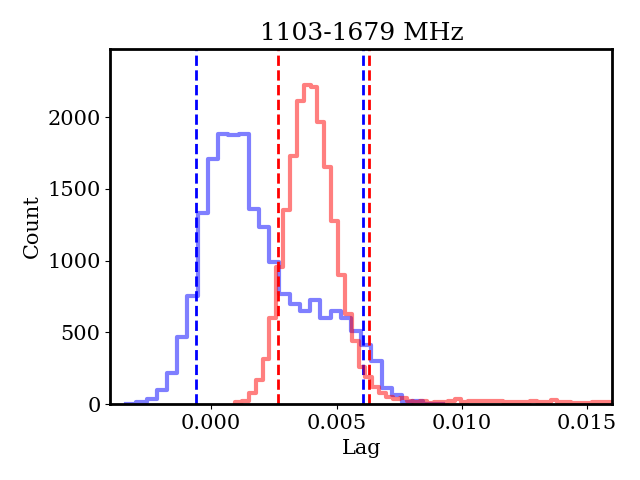}
    \includegraphics[width=0.23\textwidth]{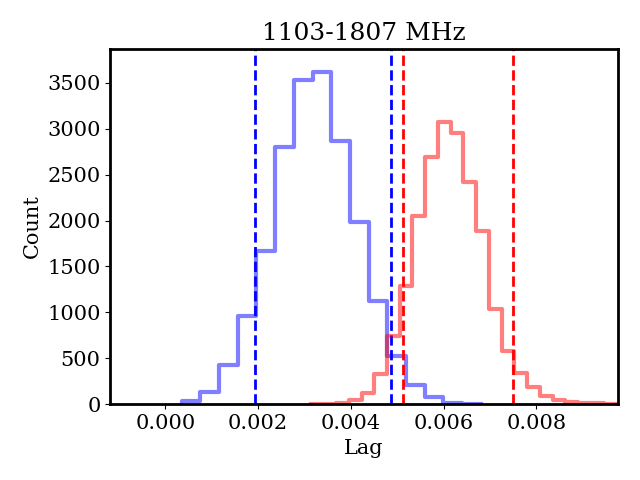}
    \includegraphics[width=0.23\textwidth]{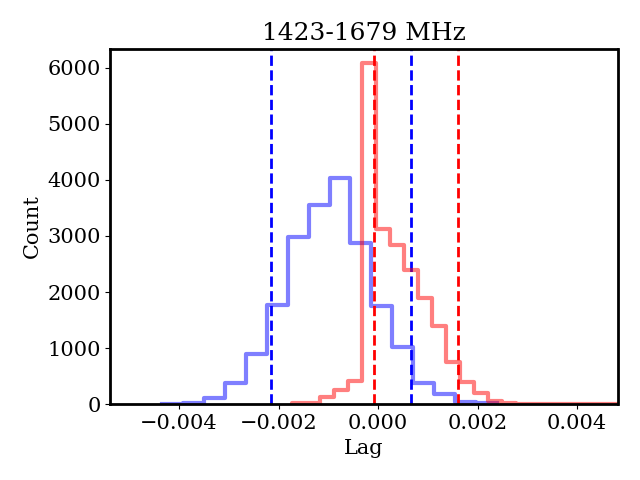}
    \includegraphics[width=0.23\textwidth]{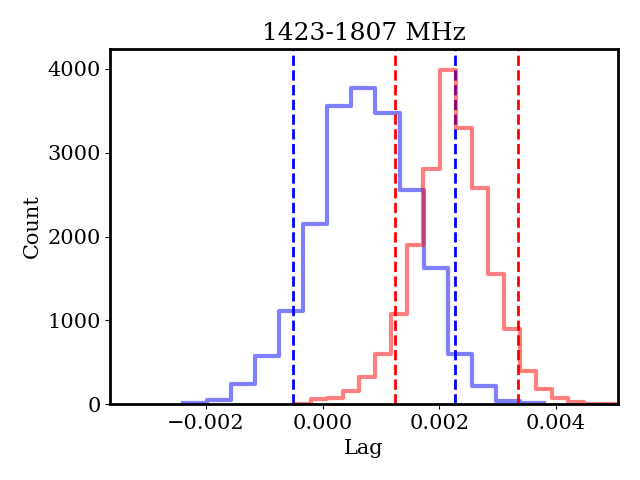}
    \includegraphics[width=0.23\textwidth]{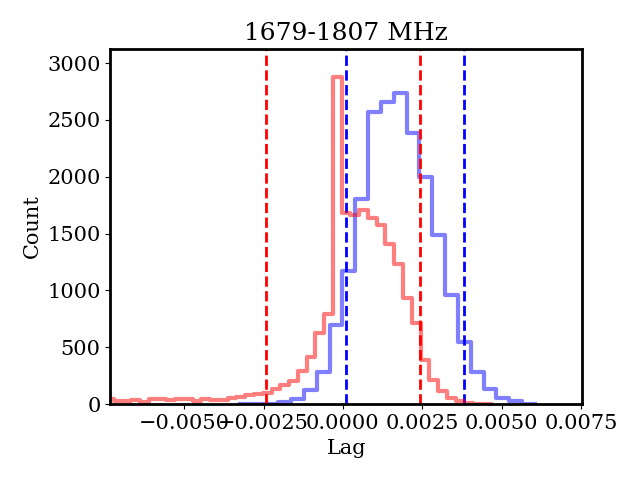}
    \caption{The lag distributions for the RCP pulses near Null 1 (red) and the LCP pulses near Null 2 (blue). The dashed vertical lines mark the 90\% confidence intervals.
    \label{fig:lag_hists}}
\end{figure*}

The histograms of the lag distributions are shown in Figure \ref{fig:lag_hists}. The procedure to obtain the lags are described in \S\ref{sec:finding_lags}.

\bibliography{das_apj}{}
\bibliographystyle{aasjournal}



\end{document}